\documentclass[aps,prd,amsmath,floatfix
,twocolumn
,superscriptaddress,nofootinbib,showpacs]{revtex4-1}

\usepackage{graphicx} 
\usepackage{amsfonts} \usepackage{subfigure} \usepackage{xspace}
\usepackage[usenames,dvipsnames]{color} \usepackage{dcolumn}
\usepackage{bm} \usepackage{hyperref} \usepackage{mathrsfs}
\usepackage[]{amsmath,amssymb} \usepackage[]{amsthm}
\usepackage{verbatim} \usepackage{appendix}
\usepackage[colorinlistoftodos]{todonotes}

\usepackage{ulem} 

\begin{document}

\title[Binary Neutron Stars with Arbitrary Spins in Numerical
  Relativity]{Binary Neutron Stars with Arbitrary Spins in Numerical
  Relativity}

\newcommand{\AEI}{\affiliation{Max Planck Institute for Gravitational Physics
(Albert Einstein Institute), Am M\"uhlenberg 1, Potsdam-Golm, 14476, Germany}} %
\newcommand{\Caltech}{\affiliation{Theoretical Astrophysics 350-17,
    California Institute of Technology, Pasadena, CA 91125, USA}}
\newcommand{\CITA}{\address{Canadian Institute for Theoretical
    Astrophysics, University of Toronto, 60~St.~George Street,
    Toronto, Ontario M5S 3H8, Canada}} %
\newcommand{\CIFAR}{\affiliation{Canadian Institute for Advanced
    Research, 180 Dundas St.~West, Toronto, ON M5G 1Z8, Canada}} %
\newcommand{\Cornell}{\affiliation{Center for Radiophysics and Space
    Research, Cornell University, Ithaca, New York 14853, USA}}
\newcommand{\DAA}{\affiliation{Department of Astronomy and
    Astrophysics, 50 St.\ George Street, University of Toronto,
    Toronto, ON M5S 3H4, Canada}}
\newcommand{\LBL}{\affiliation{Lawrence Berkeley National Laboratory,
    1 Cyclotron Rd, Berkeley, CA 94720, USA; Einstein Fellow}}
\newcommand{\WSU}{\affiliation{
Department of Physics \& Astronomy, Washington State University, Pullman, Washington 99164, USA}}
\author{Nick Tacik}\CITA\DAA
\author{Francois Foucart}\CITA\LBL
\author{Harald P. Pfeiffer}\CITA\CIFAR
\author{Roland Haas}\Caltech\AEI
\author{Serguei Ossokine}\CITA\DAA
\author{Jeff Kaplan}\Caltech
\author{Curran Muhlberger}\Cornell
\author{Matt D. Duez}\WSU
\author{Lawrence E. Kidder}\Cornell
\author{Mark A. Scheel}\Caltech
\author{B\'{e}la Szil\'{a}gyi}\Caltech

\begin{abstract}
  We present a code to construct initial data for binary neutron star
  systems in which the stars are rotating.  
  Our code, based on a formalism
  developed by Tichy, allows for arbitrary rotation axes of the
  neutron stars and is able to achieve rotation rates near rotational
  breakup.  We compute the neutron star angular momentum through quasi-local angular momentum
    integrals. When constructing irrotational binary neutron stars, we find a very small residual dimensionless spin of $\sim 2\times 10^{-4}$.  Evolutions of rotating neutron star binaries show that the magnitude of the
    stars' angular momentum is conserved, and that the spin- and
    orbit-precession of the stars is well described by
    post-Newtonian approximation.  We demonstrate that orbital
  eccentricity of the binary neutron stars can be controlled to
  $\sim 0.1\%$.  The neutron stars show
    quasi-normal mode oscillations at an amplitude which increases
    with the rotation rate of the stars.
\end{abstract}


\maketitle

\section{Introduction}

Several known binary neutron star (BNS) systems  will merge within a
Hubble time due to inspiral driven by gravitational
radiation~\cite{Lorimer2008}, most notably the Hulse-Taylor
pulsar~\cite{Hulse:1975uf}.  Therefore, binary neutron stars
constitute one of the prime targets for upcoming gravitational wave
detectors like Advanced LIGO~\cite{aLIGO1,aLIGO2} and Advanced
Virgo~\cite{AdV,TheVirgo:2014hva}.
The neutron stars in known binary pulsars have fairly long rotation
periods~\cite{Lorimer2008}.  The system J0737-3039~\cite{Lyne:2004cj}
contains the fastest known spinning neutron star in a binary with a
rotation period of 22.7ms.  This system will merge within $\sim
10^8$~years through gravitational wave driven inspiral. Globular
clusters contain a significant fraction of all known milli-second
pulsars~\cite{Lorimer2008}, which through dynamic interactions, may
form binaries~\cite{2010ApJ...720..953L,Benacquista:2011kv}. Gravitational wave driven
inspiral reduces~\cite{PetersMathews1963,Peters1964} the initially
high eccentricity of dynamical capture binaries.
Given the presence of milli-second pulsars in
globular clusters, dynamically formed BNS may contain very rapidly
spinning neutron stars with essentially arbitrary spin orientations.
Presence of spin in BNS systems does influence the evolution of the
binary.  For instance, in order to avoid a loss in sensitivity in GW
searches, one needs to account for the NS spin~\cite{Brown:2012qf}.
Furthermore, early BNS simulations~\cite{Shibata00b} of irrotational
and corotational BNS systems found that the spin
of corotating BNS noticeably increased the size of accretion discs
occurring during the merger of the two NS.  The properties of
accretion discs and unbound ejecta are intimately linked to
electromagnetic and neutrino emission from merging compact object
binaries~{\cite{metzger:11}. 
Understanding the behavior of rotating BNS systems is therefore important to quantify the expected observational signatures from such systems.
These considerations motivated a recent interest in the numerical
modeling of rotating binary neutron star systems during their last
orbits and coalescence.  Baumgarte and
Shapiro~\cite{Baumgarte:2009fw}, Tichy~\cite{Tichy:2011gw}, and East
et al~\cite{East:2012zn} presented formalisms for constructing BNS
initial data for spinning neutron stars.  Tichy proceeded to construct
rotating BNS initial data~\cite{Tichy:2012rp}; and
Ref.~\cite{Bernuzzi:2013rza} studies short inspirals and mergers of
BNS with rotation rates consistent with known binary neutron stars
(i.e. a dimensionless angular momentum of each star
$\chi=S/M^2\lesssim 0.05$), and rotation axes aligned with the orbital
angular momentum.  Very recently, Dietrich et al.~\cite{Dietrich:2015pxa} presented a comprehensive study of BNS, including a simulation of a precessing, merging BNS.
East et. al.~\cite{East:2015yea} investigate
interactions of rotating neutron stars on highly eccentric orbit.
Kastaun et al.~\cite{Kastaun:2013mv} determine the maximum spin of the
black hole remnant formed by the merger of two aligned spin rotating
neutron stars.  Tsatsin and Marronetti~\cite{Tsatsin:2013jca} present
initial data and evolutions for non-spinning, spin-aligned and
anti-aligned data sets.  

Previous studies differ in the type of initial data used:
Refs.~\cite{Baumgarte:2009fw,Tichy:2011gw,Tichy:2012rp,Bernuzzi:2013rza}
construct and utilize constraint-satisfying initial data, which also
incorporates quasi-equilibrium of the binary system.
Refs.~\cite{East:2012zn,East:2015yea} construct constraint-satisfying
data based on individual TOV stars, without regard of preserving
quasi-equilibrium in the resulting binary, but providing greatly
enhanced flexibility in the type of configurations that can be
studied, e.g. hyperbolic encounters.
Refs.~\cite{Kastaun:2013mv,Tsatsin:2013jca}, finally, only
approximately satisfy the constraint equations.
Previous studies also
differ in the rigor with which the neutron star angular momentum is
measured.  Ref.~\cite{Tichy:2012rp} merely discusses the neutron stars
based on a rotational velocity $\omega^i$ entering the initial data
formalism (cf. our Eq.~(\ref{eq:UniformRotation}) below), whereas
Refs.~\cite{Bernuzzi:2013rza,Kastaun:2013mv,East:2015yea} estimate the
initial neutron star spin either based on single star models or based
on the differences in binary neutron star initial data sets with and
without rotation, and thus neglecting the impact of interactions in
the binary.  All these studies measure the neutron star angular
momentum in the initial data.  Changes in the neutron star angular
momentum that could happen during initial
relaxation of the binary or during the subsequent evolution of the
binary are not monitored.

In this paper we study the construction of rotating binary neutron
star initial data and the evolution through the inspiral phase.  We
implement the constant rotational velocity (CRV) formalism developed
by Tichy~\cite{Tichy:2012rp}, and construct constraint satisfying BNS
initial data sets with a wide variety of spin rates, as well as different spin
\emph{directions}.  We apply quasi-local angular momentum techniques
developed for black holes to our BNS initial data sets; the
quasi-local spin indicates that we are able to construct BNS with
dimensionless angular momentum exceeding 0.4.  Evolving some of the
constructed initial data sets through the inspiral phase, we
demonstrate that we can control and reduce orbital eccentricity by an
iterative adjustment of initial data parameters controlling orbital
frequency and radial velocity of the stars, both for non-precessing
(i.e. aligned-spin binaries) and precessing binaries.  When monitoring
the quasi-angular momentum of the neutron stars during the inspiral,
we find that its magnitude is conserved, and the spin-direction
precesses in a manner consistent with post-Newtonian predictions.

This paper is organized as follows.  Section~\ref{sec:Methodology}
describes the initial data formalism and our numerical code to solve
for rotating BNS initial data.  In Sec.~\ref{sec:ID} we use this
code to study a range of initial configurations, 
with a special emphasis on the behavior of
the quasi-local spin diagnostic.  We evolve rotating BNS in
Sec.~\ref{sec:EvolutionResults}, including a discussion of
eccentricity removal, the behavior of the quasi-local spin
diagnostics, and a comparison of the precession dynamics to
post-Newtonian theory.  A discussion concludes the paper in
Sec.~\ref{sec:Discussion}. In this paper,  we work in units where $G=c=M_{\odot}=1$.

\section{Methodology}
\label{sec:Methodology}

\subsection{Formalism for irrotational binaries}
\label{sec:IrrotFormalism}

To start, we will review a formalism commonly used for the 
construction of initial data for system of irrotational binary neutron stars. 
We will then discuss how to
build upon this formalism to construct initial data for neutron stars
with arbitrary spins.

We begin with the 3+1 decomposition of the space-time metric
(see~\cite{2007gr.qc.....3035G} for a review),
\begin{equation}
ds^2 = -\alpha^2 dt^2 + \gamma_{ij}\left(dx^i +
\beta^idt\right)\left(dx^j + \beta^jdt\right).
\end{equation}

Here, $\alpha$ is the lapse function, $\beta^i$ is the shift vector
and $\gamma_{ij}$ is the 3-metric induced on a hypersurface
$\Sigma(t)$ of constant coordinate time $t$. In this decomposition, the unit normal
vector $n^{\mu}$ to $\Sigma(t)$ and the tangent vector $t^{\mu}$ to
the coordinate line $t$ are related by
\begin{equation}
t^{\mu} = \alpha n^{\mu} + \beta^{\mu},
\end{equation}
with $n_\mu=(-\alpha,0,0,0)$ and $\beta^\mu=(0,\beta^i)$.  The
extrinsic curvature of $\Sigma(t)$ is the symmetric tensor defined as
\begin{equation}
K_{\mu\nu} = -\nabla_\nu n_\mu -n_\nu \gamma^\lambda_{\phantom{\lambda}\mu} \nabla_\lambda (\ln \alpha) =
-\frac{1}{2}\mathcal{L}_n \gamma_{\mu\nu},
\end{equation}
where $\gamma_{\mu \nu}=g_{\mu \nu} + n_\mu n_\nu$ is the extension of
the 3-metric $\gamma_{ij}$ to the 4-dimensional spacetime, and $g_{\mu
  \nu}$ is the 4-metric of that spacetime. By construction, $K^{\mu
  \nu}n_\mu =0$ and we can restrict $K^{\mu \nu}$ to the 3-dimensional
tensor $K^{ij}$ defined on $\Sigma \times \Sigma$. The extrinsic
curvature $K^{ij}$ is then divided into its trace $K$ and trace-free
part $A^{ij}$:
\begin{equation}
K^{ij} = A^{ij} + \frac{1}{3}\gamma^{ij}K.
\end{equation}

We treat the matter as a perfect fluid with stress-energy tensor
\begin{equation}
T_{\mu\nu} = \left(\rho+P\right)u_{\mu}u_{\nu} + Pg_{\mu\nu},
\end{equation}
where $\rho=\rho_0 (1+\epsilon)$ is the energy density, $\rho_0$ the
baryon density, $\epsilon$ the specific internal energy, $P$ the
pressure, and $u_\mu$ the fluid's 4-velocity. For the initial value
problem, it is often convenient to consider the following projections
of the stress tensor:
\begin{eqnarray}
E &=& T^{\mu\nu}n_{\mu}n_{\nu}, \\ S &=&
\gamma^{ij}\gamma_{i\mu}\gamma_{j\nu}T^{\mu\nu},\\ J^{i} &=&
-\gamma^{i}_{\phantom{i}\mu}T^{\mu\nu}n_{\nu}.
\end{eqnarray}

We then further decompose the metric according to the conformal
transformation
\begin{equation}
\gamma_{ij} = \Psi^4\tilde{\gamma}_{ij}.
\end{equation}
Other quantities have the following conformal
transformations:
\begin{eqnarray}
E &=& \Psi^{-6}\tilde{E}, \\ S &=& \Psi^{-6}\tilde{S}, \\ J^{i} &=&
\Psi^{-6}\tilde{J}^i, \\ A^{ij} &=& \Psi^{-10}\tilde{A}^{ij},\\ \alpha
&=& \Psi^{6}\tilde{\alpha}.
\end{eqnarray}

$\tilde{A}^{ij}$ is related to the shift and the time derivative of the conformal
metric, $\tilde{u}_{ij}=\partial_t\tilde{\gamma}_{ij}$ by
\begin{equation}
\tilde{A}^{ij} =
\frac{1}{2\tilde{\alpha}}\left[\left(\tilde{\mathbb{L}}\beta\right)^{ij}-\tilde{u}^{ij}\right],
\end{equation}
where $\tilde{\mathbb{L}}$ is the conformal longitudinal operator whose action on a vector $V^i$ is
\begin{equation}
\left(\tilde{\mathbb{L}}V\right)^{ij} = \tilde{\nabla}^iV^j +
\tilde{\nabla}^jV^i -
\frac{2}{3}\tilde{\gamma}^{ij}\tilde{\nabla}_kV^k,
\end{equation}
and $\tilde{\nabla}$ is the covariant derivative defined with respect to 
the conformal 3-metric $\tilde \gamma_{ij}$.

In the 3+1 formalism, the Einstein equations are decomposed into a set
of evolution equations for the metric variables as a function of $t$,
and a set of constraint equations on each hypersurface
$\Sigma(t)$. The initial data problem consists in providing quantities
$g_{\mu \nu}(t_0)$ and $K_{\mu \nu}(t_0)$ which satisfy the
constraints on $\Sigma(t_0)$ and represent initial conditions with the
desired physical properties (e.g. masses and spins of the objects,
initial orbital frequency, eccentricity, etc.).We solve the constraint
equations using the Extended Conformal Thin Sandwich (XCTS)
formalism~\cite{York1999}, in which the constraints take the form of
five nonlinear coupled elliptic equations.  The XCTS equations can be
written as

\begin{eqnarray}
2\tilde{\alpha}\bigg[\tilde{\nabla}_j\left(\frac{1}{2\tilde{\alpha}}\big(\tilde{L}\beta\big)^{ij}\right)-\tilde{\nabla}_j\left(\frac{1}{2\tilde{\alpha}}\tilde{u}^{ij}\right) && \nonumber\\
\label{eq:XCTS-Shift}
-\frac{2}{3}\Psi^6\tilde{\nabla}^iK-8\pi\Psi^4\tilde{J}^i\bigg] &=&0,
\end{eqnarray}

\begin{eqnarray}
\tilde{\nabla}^2\Psi - \frac{1}{8}\Psi\tilde{R} -
\frac{1}{12}\Psi^5K^2  \qquad\quad && \nonumber \\
\label{eq:XCTS-ConformalFactor}
+\frac{1}{8}\Psi^{-7}\tilde{A}_{ij}\tilde{A}^{ij} +
2\pi\Psi^{-1}\tilde{E} &=& 0,
\end{eqnarray}

\begin{eqnarray}
&&\tilde{\nabla}^2\left(\tilde{\alpha}\Psi^7\right) -
\left(\tilde{\alpha}\Psi^7\right)\bigg[\frac{1}{8}\tilde{R}+\frac{5}{12}\Psi^4K^2+\frac{7}{8}\Psi^{-8}\tilde{A}_{ij}\tilde{A}^{ij}\nonumber \\
\label{eq:XCTS-Lapse}
&&+2\pi\Psi^{-2}\big(\tilde{E}+2\tilde{S}\big)\bigg]=-\Psi^5\left(\partial_{t}K
- \beta^{k}\partial_kK\right).
\end{eqnarray}

We solve these equations for the conformal factor $\Psi$, the
densitized lapse $\tilde\alpha\Psi^7$ and the shift $\beta^i$. $\tilde{E}$,
$\tilde{S}$ and $\tilde{J}^i$ determine the matter content of the
slice. The variables $\tilde{\gamma}_{ij}$, $\tilde{u}_{ij} =
\partial_t\tilde{\gamma}_{ij}$, $K$ and $\partial_t K$ are freely
chosen.

  If we work in a coordinate system corotating with the binary,
  $\tilde{u}_{ij} = 0$ and $\partial_t K = 0$ are natural choices for
  a quasi-equilibrium configuration. Following earlier work~\cite{Taniguchi2007,TaniguchiEtAl:2006,FoucartEtAl:2008}, we also choose to
  use maximal slicing, $K=0$, and a conformally flat metric,
  $\tilde{\gamma}_{ij}=\delta_{ij}$.  
  Maximal slicing is a gauge choice that determines the
    location of the initial data hypersurface in the embedding space
    time.  Conformal flatness is used for computational convenience;
    rotating black holes are known to be not conformally
    flat~\cite{GaratPrice:2000}, and so this simplifying assumption should be revisited in the future.

In addition to solving these equations for the metric variables, we
must impose some restrictions on the matter. In particular, the stars
should be in a state of approximate hydrostatic equilibrium in the comoving
frame. 
This involves solving the Euler equation and the continuity
equation. For an irrotational binary, the first integral of the Euler
equation leads to the condition
\begin{equation}
h\alpha\frac{\gamma}{\gamma_0} = C,
\label{eq:Euler0}
\end{equation}
where $C$ is a constant, hereafter referred to as the Euler constant, the enthalpy $h$ is defined as
\begin{equation}
h = 1+\epsilon + \frac{P}{\rho_0},
\end{equation}
and we have introduced
\begin{eqnarray}
\gamma &=& \gamma_n\gamma_0\left(1 -
\gamma_{ij}U^iU^j_0\right),\\ \gamma_0 &=& \left(1 -
\gamma_{ij}U^i_0U^j_0\right)^{-1/2},\\ \gamma_n &=& \left(1 -
\gamma_{ij}U^iU^j\right)^{-1/2},\\ U^i_0 &=& \frac{\beta^{i}}{\alpha}.
\end{eqnarray}
The 3-velocity $U^i$ is defined by
\begin{eqnarray}
u^{\mu} &=& \gamma_n (n^\mu + U^\mu),\\ U^\mu n_\mu &=& 0.
\end{eqnarray}

The choice of $U^i$, which is unconstrained in this formalism, is an
important component is determining the initial conditions in the
neutron star. For irrotational binaries (non-spinning neutron stars),
there exists a potential $\phi$ such that
\begin{equation}
U^i = \frac{\Psi^{-4}\tilde{\gamma}^{ij}}{h\gamma_n}\partial_j\phi.
\end{equation}
The continuity equation can then be written as a second-order elliptic
equation for $\phi$:
\begin{equation}
\label{eq:Continuity}
\frac{\rho_0}{h}\nabla^{\mu}\nabla_{\mu}\phi +
\left(\nabla^{\mu}\phi\right)\nabla_{\mu}\frac{\rho_0}{h}=0.
\end{equation}

Under the assumption of the existence of an approximate helicoidal Killing vector $\xi$~\cite{Teukolsky:1998sh,Shibata:1998um}, this equation becomes
\begin{align}
\rho_0\,\bigg\{\!\!&-\tilde{\gamma}^{ij}\partial_i\partial_j\phi + 
\frac{h\beta^{i}\Psi^4}{\alpha}\partial_i\gamma_n+hK\gamma_n\Psi^4
\nonumber \\
&\quad+\left[\tilde{\gamma}^{ij}\tilde{\Gamma}^k_{ij}+\gamma^{ik}\partial_i\left(\ln\frac{h}{\alpha\Psi^2}\right)\right]\partial_k\phi 
\bigg\} \nonumber \\
&=
\tilde{\gamma}^{ij}\partial_i\phi\partial_j\rho_0-\frac{h\gamma_n\beta^i\Psi^4}{\alpha}\partial_i\rho_0. 
\label{eq:Continuity0}
\end{align}

Another simple choice for $U^i$ is to enforce corotation of the star,
i.e. $U^i=U^i_0$. This would be the case if neutron star binaries were
tidally locked.  However, viscous forces in neutron stars are expected
to be insufficient to impose tidal locking~\cite{BildstenCutler1992},
and the neutron star spins probably remain close to their value at
large orbital separations.

Once we have obtained $h$ from the metric and $U^i$, the other
hydrodynamical variables can be recovered if we close the system by the
choice of an equation of state for cold neutron star matter in
$\beta$-equilibrium, $P = P(\rho_0)$ and $\epsilon=\epsilon(\rho_0)$.
Throughout this work, we use a polytropic equation of state, $P=\kappa\rho_0^\Gamma$, with $\Gamma=2$. The internal energy, $\epsilon\rho_0$, satisfies 
\begin{equation}
\epsilon\rho_0=\frac{P}{\Gamma-1}.
\end{equation}

The boundary conditions of our system of equations are quite
simple. At the outer boundary of the computational domain (which we
approximate as ``infinity'' and is in practice $10^{10}M_{\odot}$), we require the metric to be Minkowski
in the inertial frame, and so in the corotating frame we have
\begin{eqnarray}
{\bm \beta} &=& {\bm \Omega}_0 \times {\bm r} + \dot a_0 {\bm
  r},\\ \alpha &=& 1,\\ \Psi &=& 1,
\end{eqnarray}
with ${\bm \Omega}_0$ the initial orbital frequency of the binary and
$\dot{a}_0$ the initial inspiral rate of the binary. We choose ${\bm
  \Omega}_0=(0,0,\Omega_0)$, with $\Omega_0$ and $\dot{a}_0$ as freely
specifiable variables that determine the initial eccentricity of the
binary.

At the surface of each star, the boundary condition can be easily
inferred from the $\rho_0=0$ limit of equation~(\ref{eq:Continuity0}):
\begin{equation}
\tilde{\gamma}^{ij}\partial_i\phi\partial_j\rho_0 =
\frac{h\gamma_n\beta^{i}\Psi^4}{\alpha}\partial_i\rho_0.
\end{equation}

Finally, we discuss how a first guess for the orbital angular velocity
$\Omega_0$ can be obtained for a non-spinning system.  The force
balance equation at the centre of the NS is
\begin{equation}\label{eq:Nablah}
\nabla\ln{h}=0.
\end{equation}
Neglecting any infall velocity, this condition guarantees that the
binary is in a circular orbit. This is only an
approximation as there is really some infall velocity, but this still
leads to low eccentricity binaries with $e\sim 0.01$. From the
integrated Euler equation, we can write this condition as
\begin{equation}
\nabla\ln{h} = \nabla\left(\ln{\frac{\gamma_0}{\alpha\gamma}}\right)=0,
\end{equation}
or, by using the definitions of $\gamma_0$, and $\gamma$,
\begin{equation}
\nabla\ln\left(\alpha^2 - \gamma_{ij}\beta^{i}\beta^{j}\right) =
-2\nabla\ln{\gamma}.
\label{eq:OmegaEq}
\end{equation}
If we decompose $\beta^i$ in its inertial component $\beta^i_0$ and
its comoving component according to
\begin{equation}
{\bm \beta} = {\bm \beta}_0 + {\bm \Omega}_0 \times {\bm r} + \dot a_0
{\bm r},
\end{equation}
this can be written as a quadratic equation for the orbital angular
velocity $\Omega_0$ (neglecting the dependence of $\gamma$ on the
orbital angular velocity $\Omega_0$).  In practice, we solve for
$\Omega_0$ by projecting Eq.~(\ref{eq:OmegaEq}) along the line
connecting the center of the two stars.\footnote{Along the other
  directions, the enthalpy $h$ is corrected so that force balance is
  enforced at the center of the star, according to the method
  described in~\cite{Foucart:2010eq}}

The exact iterative procedure followed to solve in a consistent manner
the constraint equations, the elliptic equations for $\phi$, and the
algebraic equations for $h$ (including on-the-fly computation of
$\Omega_0$ and of the constant in the first integral of Euler
equation) is detailed in Section~\ref{sec:IDalgorithm}.

Once a quasi-equilibrium solution has been obtained by this method,
lower eccentricity systems can be generated by modifying $\Omega_0$
and $\dot{a}_0$, following the methods developed by Pfeiffer et
al.~\cite{Pfeiffer-Brown-etal:2007}.

\subsection{Formalism for Spinning Binaries}
We will now discuss how to alter the formalism discussed above to
incorporate spinning BNS. Although several formalisms have been
introduced in the past \cite{Marronetti:2003gk}\cite{Baumgarte:2009fw}, we will
follow the work of Tichy (2011)\cite{Tichy:2011gw}. A
first obvious difference is that we can no longer write the velocity
solely in terms of the gradient of a potential. Following Tichy, we
break the velocity up into an irrotational part, and a new rotational
part $W$:
\begin{equation}
U^i =
\frac{\Psi^{-4}\tilde{\gamma}^{ij}}{h\gamma_n}\left(\partial_j\phi+W_j\right),
\end{equation}
where it is natural, although not required, for $W$ to be
  divergenceless.

Following the assumptions stated in Tichy\cite{Tichy:2011gw}, the continuity equation becomes
\begin{align}
\rho_0\,\bigg\{\!\!&-\tilde{\gamma}^{ij}\partial_i\big(\partial_j\phi+W_j\big)  + \frac{h\beta^i\Psi^4}{\alpha}\partial_i\gamma_n + hK\gamma_n\Psi^4 \nonumber\\
&\qquad+\Big[\tilde{\gamma}^{ij}\tilde{\Gamma}^k_{ij}+\gamma^{ik}\partial_i\big(\ln \frac{h}{\alpha\Psi^2}\big)\Big] 
\big(\partial_k\phi+W_k\big) \bigg\} \nonumber \\
&=\tilde{\gamma}^{ij}\big(\partial_i\phi+W_i\big)\partial_j\rho_0 - \frac{h\gamma_n\beta^i\Psi^4}{\alpha}\partial_i\rho_0.
\label{eq:Continuity}
\end{align}
Eq.~\ref{eq:Continuity} then is the same as in the irrotational
case, cf. Eq.~\ref{eq:Continuity0}, under the replacement $\partial_i\phi\rightarrow\partial_i\phi+W_i$.

Taking the limit $\rho_0\to 0$ in Eq.~(\ref{eq:Continuity}) yields the
boundary condition at the surface of each star:
\begin{equation}
\tilde{\gamma}^{ij}\left(\partial_i\phi+W_i\right)\partial_j\rho_0=\frac{h\gamma_n\beta^i\Psi^4}{\alpha}\partial_i\rho_0.
\end{equation}

The solution of the Euler equation is no longer as simple as it was
previously, in Eq.~\ref{eq:Euler0}. As
shown in Tichy(2011)\cite{Tichy:2011gw}, the solution is
now
\begin{equation}
h = \sqrt{L^2 -
  \left(\nabla_i\phi+W_i\right)\left(\nabla^i\phi+W^i\right)},
\end{equation}
where
\begin{equation}
L^2 =
\frac{b+\sqrt{b^2-4\alpha^4\left(\left(\nabla_i\phi+W_i\right)W^i\right)^2}}{2\alpha^2},
\end{equation}
and
\begin{equation}
b =
\left(\beta^i\nabla_i\phi+C\right)^2+2\alpha^2\left(\nabla_i\phi+w_i\right)w^i.
\end{equation}

Finally, the method discussed previously of modifying the star's
angular velocity is now no longer as simple. The equation is modified
to
 \begin{equation}
\nabla\ln\left(\alpha^2-\gamma_{ij}\beta^{i}\beta^{j}\right)=-2\nabla\ln\Gamma,
\end{equation}
where
\begin{equation}
\Gamma
=\frac{\gamma_n\left(1-\left(\beta^i+\frac{W^i\alpha}{h\gamma_n}\right)\frac{\nabla_i\phi}{\alpha
    h\gamma_n}- \frac{W_i W^i}{\alpha^2\gamma_n^2}\right) } { \sqrt{ 1
    - \left(\frac{\beta^i}{\alpha}+\frac{W^i}{h\gamma_n}\right)
    \left(\frac{\beta_i}{\alpha}+\frac{W_i}{h\gamma_n}\right) } } .
\end{equation}

Let us now discuss the choice of the spin term, $W$. This term is, in
principle, freely chosen, and so we must choose it so as to best
represent the physical situation at hand - namely a uniform rotation
with constant angular velocity. As suggested by Tichy(2011)\cite{Tichy:2011gw} and
Tichy(2012)\cite{Tichy:2012rp}, a reasonable choice for $W$ is 
\begin{equation}
\label{eq:UniformRotation}
W^i = \epsilon^{ijk}\omega^{j}r^k,
\end{equation}
where $r^k$ is the position vector centered at the star's centre,
$\omega^j$ represents an angular velocity vector and $\epsilon^{ijk}=\left\{\pm 1,0\right\}$. This leads to a vector field $W^i$ with vanishing divergence in the conformal metric $\tilde{g}_{ij}=\delta_{ij}$. Alternatively, one might prefer a vector field $V^i$ with vanishing divergence with respect to the physical metric $g_{ij}=\Psi^4\delta_{ij}$. Owing to the conformal transformation properties of the divergence operator, $V^i$ is given by
\begin{equation}
\label{eq:ConformalUniformRotation}
V^i=\Psi^{-6}W^i.
\end{equation}
Here, we generally use $W^i$ as we have found that it leads to initial data which is closer to being in equilibrium, as we will further discuss in section~\ref{sec:QNModes}.

\subsection{Solving the elliptic equations}

\begin{figure}
\includegraphics[width=0.95\columnwidth,trim=0 185 0 0, clip=true]{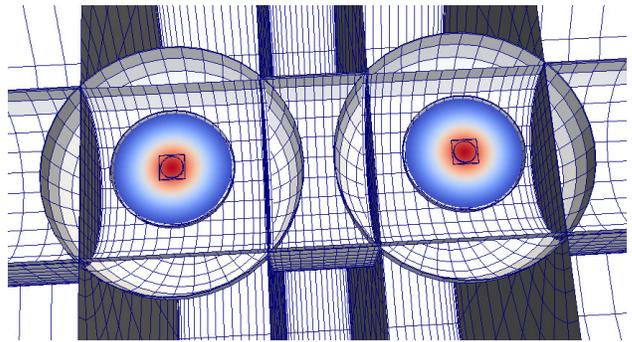}
\caption{\label{fig:TheDomain} Visualization in the x-y plane of the
  domain decomposition used in our initial data solve. The colour map
  represents the density of the stars.  }
\end{figure}

In the previous sections, we have reduced the Einstein
  constraints, Eqs.~(\ref{eq:XCTS-Shift})--(\ref{eq:XCTS-Lapse}), as
  well as the continuity equation~(\ref{eq:Continuity}) to elliptic
  equations.  We solve these equations with the multi-domain
  pseudo-spectral elliptic solver developed in~\cite{Pfeiffer2003}, as
  modified in~\cite{FoucartEtAl:2008} for matter.  The
  computational domain is subdivided into individual subdomains as
  indicated in Fig.~\ref{fig:TheDomain}: The region near the center of
  each star is covered by a cube, overlapping the cube is a spherical
  shell which covers the outer layers of the star.  The outer boundary
  of this shell is deformed to conform to the surface of the star.
  This places all surfaces at which the solution is not smooth at a
  subdomain-boundary, which preserves the exponential convergence of
  spectral methods.  
  Another spherical shell surrounds each star.  The
  inner shells representing the stars and their vicinity are embedded
  into a structure of five concentric cylinders with three rectangular
  blocks along the axis connecting the centers of the neutron stars,
  which overlap the inner spherical shells.  The cylinders/blocks in
  turn are overlapped at large radius by one further spherical shell
  centered half-way between the two neutron stars.  Using an inverse radial mapping, the outer radius of the outer sphere is placed at $10^{10}$.
 
  All variables are decomposed on sets of basis functions depending on the subdomain. 
  The resolution of each domain (i.e., the number of colocation points used) 
is chosen at the start of the initial data solve, and then
subsequently modified several times using an adaptive procedure
described below. In this paper, when discussing the total resolution of the
domain, we use the notation 
\begin{equation}
N^{1/3} = \left(\sum
N_i\right)^{1/3},
\end{equation}
with $N_i$ the number of collocation points in the $i$th subdomain. $N^{1/3}$ is thus the cube root of the total number of
collaction points in all subdomains.  

\subsection{Construction of quasi-equilibrium initial data}
\label{sec:IDalgorithm}

Construction of initial data for rotating binary neutron stars
begins with selecting the physical properties of the system: the equation of
state of nuclear matter, the coordinate separation $d$ between the
neutron stars, the baryon masses $M^b_1$ and $M^b_2$ of the two stars, and their
spin vectors ${\bm \omega}_{\rm rot,1}$ and ${\bm \omega}_{\rm
  rot,2}$. We also choose the orbital angular frequency $\Omega_0$
and the initial inspiral rate $\dot{a}_0$.  

We generally begin by setting $\Omega_0$ to the value for the orbital frequency of a similar irrotational BNS (where $\Omega_0$ is determined by the condition of quasi-circularity, Eq.~(\ref{eq:Nablah})), and $\dot a_0=0$.  These values are then adjusted following the eccentricity reduction method
developed by Pfeiffer et al.~\cite{Pfeiffer-Brown-etal:2007}. Finally, we use a flat conformal metric, $\tilde\gamma_{ij}=\delta_{ij}$,
and maximal slicing, $K=0$.

Once all these quantities are fixed, we need to solve self-consistently Eqs.~(\ref{eq:XCTS-Shift})--(\ref{eq:XCTS-Lapse}) for the Einstein-constraints, the continuity equation Eq.~(\ref{eq:Continuity}), while simultaneously satisfying conditions to enforce the desired masses of the stars. 
To do so, we follow an iterative procedure developed originally for black hole-neutron star
binaries~\cite{FoucartEtAl:2011}.

First, we choose initial guesses
for the conformal metric and hydrodynamical variables, using an
analytical superposition of two isolated boosted neutron stars.

We then obtain constraint-satisfying initial conditions by applying
the following iterative procedure, where $n$ represents the iteration number:
\begin{enumerate}
\item \label{step:1} 
  Solve the nonlinear XCTS system for the set of metric variables $X=(\beta^i,\Psi,\alpha \Psi)$,
  assuming fixed values of the conformal source terms
  $(\tilde{E},\tilde{S},\tilde{J}^i$). The new value $X^{n+1}$ of the
  metric variables is obtained from their old value $X^n$ and,
  following the relaxation scheme used in \cite{FoucartEtAl:2008}, the
  solution of the XCTS equations $X^*$, using
\begin{equation}
\label{eq:Relaxation}
X^{n+1}=0.3X^*+0.7X^n.
\end{equation}
\item Locate the surface of each star.  Representing the surface in polar coordinates centered on each star as $R_s^n(\theta,\phi)$, we determine $R_s^n$ to satisfy~\cite{FoucartEtAl:2008} $h(R^n_{\rm
  s}(\theta,\phi),\theta,\phi)=1$.
To ensures that the grid-boundary $R_b$ converges to the surface of the star, we occassionally modify the numerical grid such that $R_b(\theta,\phi)=R^n_s(\theta,\phi)$. Because this requires a re-initialization of the elliptic solver, the grid is only modified if the stellar surface has settled down, specifically, if
\begin{equation}
||R^n_{\rm s} - R^{n-1}_{\rm s}|| < 0.1 ||R^n_{\rm s} - R_b||.
\end{equation}
Here $||\;.\;||_2$ denotes the L2-norm over the surface.

\item For each neutron star, fix the constant in Euler's first
  integral so that the baryon mass of the neutron star matches the desired
  value.  We compute the baryon mass as a function of the Euler constant $C$
through
\begin{equation}
M^b_{\rm NS} = \int_{\rm
  NS}\rho_0\Psi^6\sqrt{\frac{1}{1-\gamma_{ij}U^iU^j}}dV,
\end{equation}
and utilize the secant method
to drive the mass to the desired value.
\item If desired, adjust the orbital frequency to ensure force-balance at the
  center of each star by solving Eq.~(\ref{eq:OmegaEq}). This step is
  skipped if the orbital frequency is fixed through iterative eccentricity removal, cf. Sec.~\ref{sec:EccRemoval}.
\item Solve the elliptic equation for the velocity potential $\phi$,
  and obtain the next guess for $\phi$ using the same relaxation
  method shown in Eq.~\ref{eq:Relaxation}.
\item  Check whether all equations are satisfied to the desired accuracy. If yes, proceed. If no, return to Step~\ref{step:1}.
\item Compute the truncation error of the current solution by
    examining the spectral expansion of the XCTS variables. If this
    truncation error is undesirably large (typically, if it is
    $>10^{-9}$), then adjust the number of grid-points in the
    domain-decomposition and return to Step~\ref{step:1}. The adjustment
is based on the desired target truncation eror and the measured convergence rate of the solution, cf.~\cite{Szilagyi:2014fna}.
\end{enumerate}

\subsection{Quasi-Local Angular Momentum}
\label{sec:QLSpinExplanation}

The goal of the present paper is to construct spinning BNS initial data and to evolve it.
Therefore, we need diagnostics to measure the NS spin, for which we use techniques originally developed for black holes.
It is common to discuss the spins of black holes in terms of their
dimensionless spin $\chi$,
\begin{equation}\label{eq:chi}
\chi = \frac{S}{M^2}.
\end{equation}
Here, $S$ is the angular momentum of the black hole, and $M$ is its
Christodoulou mass~\cite{Christodoulou70},
\begin{equation}\label{eq:ChristodoulouMass}
M^2 = M_{\rm irr}^2 + \frac{S^2}{4M_{\rm irr}^2}.
\end{equation}
The irreducible mass $M_{\rm irr}$ is defined based on the area of the
hole's apparent horizon, $M_{\rm irr}=\sqrt{A/16\pi}$. The angular
momentum is computed with a surface integral over the apparent
horizon~\cite{BrownYork1993,Ashtekar2001,Ashtekar2003},
\begin{equation}\label{eq:QLspin}
S= \frac{1}{8\pi}\oint_{\mathcal{H}}\phi^is^jK_{ij}dA
\end{equation}
where $\mathcal{H}$ is the black hole's apparent horizon, $s^j$ is the
outward-pointing unit-normal to $\mathcal{H}$ within the $t={\rm const}$
hypersurface, and $\phi^i$ is an azimuthal vector field tangent to
$\mathcal{H}$.  For spacetimes with axisymmetry, $\phi^i$ should be
chosen as the rotational Killing vector.  In spacetimes without an
exact rotational symmetry (e.g. the spacetime of a binary black hole
system), one substitutes an \emph{approximate Killing
  vector}\cite{Cook2007,Lovelace2008} (AKV).  Ref.~\cite{Lovelace2008}
introduces a minimization principle to define $\phi^i$, resulting in
an Eigenvalue problem. The three eigenvectors with the lowest
eigenvalues (i.e. smallest shear) are taken and used to compute
the three components of the spin.

In this paper, we explore the application of quasi-local spin measures
to neutron stars.  In the absence of apparent horizons $\mathcal{H}$,
we need to choose different surface(s) to evaluate Eq.~(\ref{eq:QLspin}). 

When constructing initial data, the stellar surface ${\cal S}$ is
already determined, so one obvious choice is to integrate over the
stellar surface ${\cal S}$.  To estimate the ambiguity in quasi-local
spin, we furthermore compute $S$ by integrating over coordinate
spheres with radii ranging from just outside ${\cal S}$ to larger by
about $70\%$.  During the evolution, the stars change shape and
may even loose mass in tidal tails.  Because of these complications,
the SpEC evolution code does not track the location of the
stellar surface during the evolution, and we shall only
monitor $S$ on coordinate spheres.

It is useful to compute a dimensionless spin $\chi$, for instance, for
post-Newtonian comparisons.  In the absence of a horizon,
Eq.~(\ref{eq:ChristodoulouMass}) is meaningless and we need a different
choice for the mass-normalization.  Instead, we normalize by each 
star's ADM mass, $M_{\rm{ADM}}$, i.e. 
\begin{equation}
\chi\equiv \frac{S}{M_{\rm ADM}^2}.
\end{equation}
The ADM mass is determined by computing the ADM mass of an equilibrium configuration of a single uniformly rotating polytrope in isolation with the same baryon mass and angular momentum as those measured in our binary systems.

The results of the
quasi-local spin measures are described in
section~\ref{sec:QLSpinProperties}, which shows that this procedure is
numerically robust.

Finally, let us discuss, from an order of magnitude perspective, how
the star's dimensionless spin is related to its more commonly used physical properties.
We start with the Newtonian relation $S=2\pi I/P$ between angular
momentum $S$, moment of inertia $I$, and rotational period $P$.
Writing further $I=f\,R^2\,M$, with the dimensionless constant $f$
depending on the stellar density profile, we have 
\begin{align}
\chi&\sim\frac{2\pi c}{G}\frac{fR^2}{PM} \nonumber\\
&=
0.48\,\Big(\frac{f}{0.33}\Big)\Big(\frac{R}{12{\rm
    km}}\Big)^{\!2}\Big(\frac{M}{1.4M_{\odot}}\Big)^{\!-1}\Big(\frac{P}{1{\rm
    ms}}\Big)^{\!-1}\!\!.
\end{align}
The factor $c/G$ arises from the transition to geometric units. 

This --quite simplistic-- estimate shows that millisecond pulsars will
have appreciable dimensionless spin $\chi$.  Centrifugal breakup of
rapidly rotating neutron stars happens at a dimensionless spin in the
range $0.65-0.70$~\cite{Lo:2010bj}, with only small dependence on the
equation of state and neutron star mass.  Ansorg et
al~\cite{Ansorg:2003br} studied in detail $\Gamma = 2$ polytropes, the
equation of state we use here.  Ref.~\cite{Ansorg:2003br} finds a
dimensionless spin at mass-shedding of $\chi=0.57$.

\section{Initial Data Results}
\label{sec:ID}

In this section, we will demonstrate that our code can robustly
construct contraint-satisfying initial data for BNS systems with
arbitrary spins. As discussed in section~\ref{sec:IDalgorithm}, our code consists of a
solver that runs for a number of iterations at constant
resolution, and then the resolution is increased and this process
restarts. We will therefore demonstrate that appropriate quantities
converge with both the iterations of iterative scheme described above in
Section~\ref{sec:IDalgorithm} and with
resolution as the resolution increases.

\begin{table*}
\begin{tabular}{l|cc|ccc|cc}
Name & $M^b_{\rm NS}$ & ${\bm \omega}$ & $D_0$ & $\Omega_0 \times 10^{3}$ & $\dot{a}_0 
\times 10^{5}$  & $M_{\rm ADM}$ &  $\vec\chi$ 
\\\hline
{\tt S.4z} & 1.7745 & $0.01525\hat{z}$ & 47.2 & 5.09594 & -1.75 & 1.648 & $0.3765\hat{z}$ \\
{\tt S-.05z} & 1.7745 & $-0.00273\hat{z}$ & 47.2 & 5.11769 & -1.71 & 1.640 & $-0.05018\hat{z}$ \\
{\tt S.4x} & 1.7745 & $0.01525\hat{x}$ & 47.2 & 5.10064 & -2.36 & 1.648 & $0.3714\hat{x}$\\
\end{tabular}
\caption{\label{tab:InitialData}
Parameters for the initial data sets used in testing the initial data solver: $M^b_{\rm NS}$ and $\omega^i$ are baryon mass and rotational parameter for either neutron star (the same values are used); $D_0$, $\Omega_0$ and $\dot a_{0}$ represent coordinate separation between the centers of the stars, the orbital frequency, and the radial expansion; $\vec\chi$ is the dimensionless spin vector computed from the initial data set. In each case we use a polytropic equation of state, $P=\kappa\rho_0^{\Gamma}$, with $\Gamma=2$ and $\kappa=123.6$.}
\end{table*}

\subsection{Convergence of the Iterative Procedure}

At each step of the iterative procedure, the Euler constant of each
star is modified to achieve a desired stellar baryon mass, based on
the current matter distribution inside the star.  We expect that the Euler constant converges during the iterations at a fixed resolution. Figure~\ref{fig:EulerConstConvergence}  shows the behavior of the Euler Constant during iterations at the lowest initial data resolution, R0.  
We show three runs of interest, one
with large aligned spins ({\tt S.4z}), one with large precessing spin ({\tt S.4x}), and one with small anti-aligned spins ({\tt S-.05z}). The properties of these configurations are shown in table~\ref{tab:InitialData}.
In all three cases we see agreement between neighboring iterations at the
$10^{-5}-10^{-6}$ level by the end of iterating at this resolution. At the highest resolutions, these
differences are down to, typically, the $10^{-9}-10^{-10}$ level.
This can be compared to Fig.~3 of \cite{GourgoulhonEtAl2001a}.
Although not shown here, other free quantities converge similarly to the Euler constant.

\begin{figure}
\includegraphics[width=0.95\columnwidth]{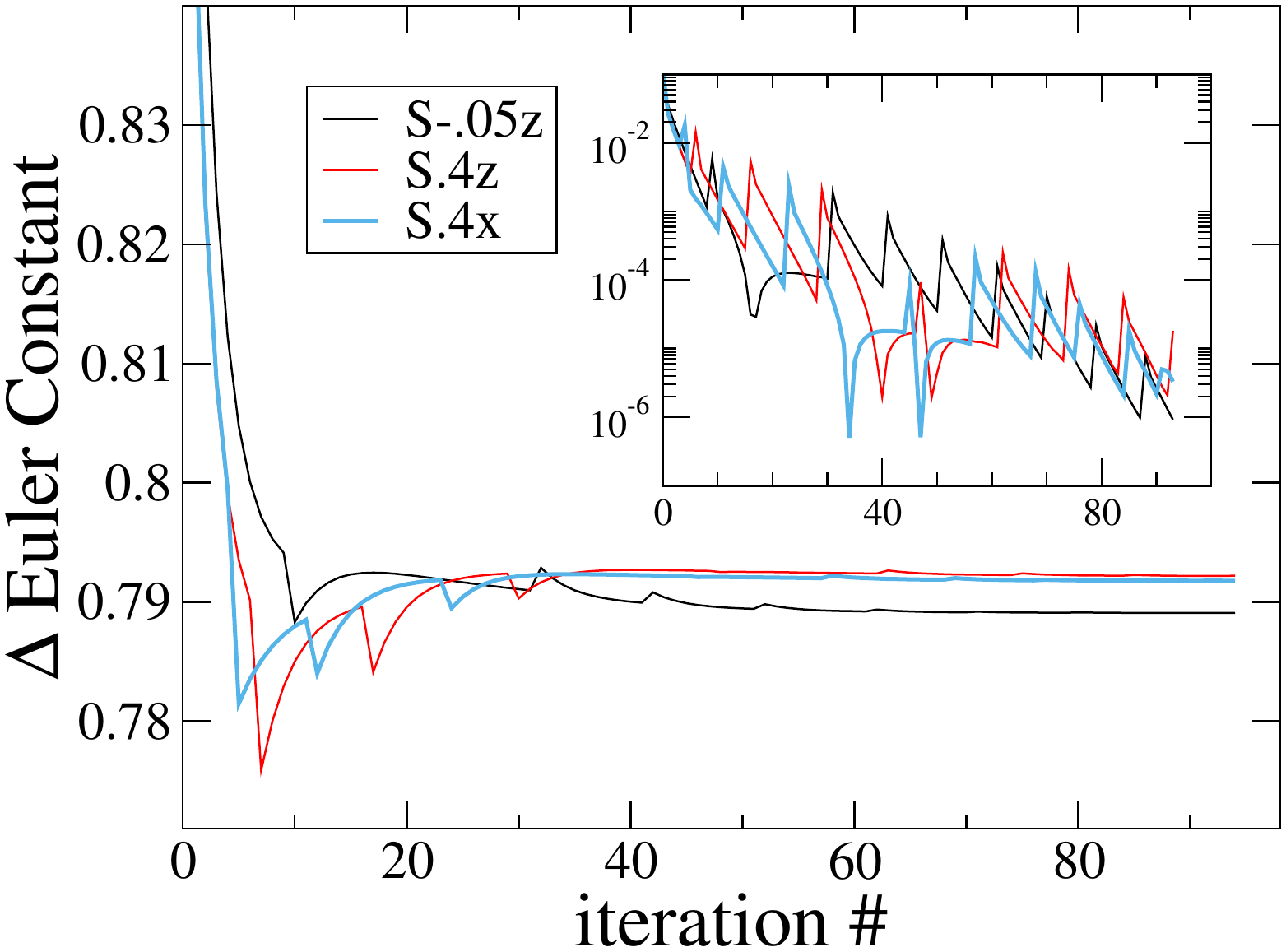}
\caption{\label{fig:EulerConstConvergence}
Convergence of the Euler
  constant during iteration at the lowest resolution R0. The inset shows the difference
  between values at subsequent iterations.}
\end{figure}

\subsection{Convergence of the Solution}

Having established that our iterative procedure converges as intended,
we now turn out attention to the convergence of the solution with
resolution. 
To demonstrate it, we will look at the Hamiltonian and
momentum constraints, and the differences between measured physical
quantities - the ADM energy and ADM angular momentum, and the surface
fitting coefficients of the stars. As our initial data representation
is fully spectral, we expect that these quantities should converge
exponentially with resolution. Note that when we discuss the value of
a quantity at a certain resolution, we are referring to the value of
that quantity after the final iterative step at that
resolution.  

Figure~\ref{fig:HamMom} shows the convergence of the Hamiltonian
constraint and the Momentum constraint for our three runs of
interest.  These are computed during the last iterative solve at each
resolution.  The data plotted are computed as
\begin{equation}
H = ||\frac{R_\Psi}{8\Psi^5}||,
\end{equation}
\begin{equation}
M = ||\frac{R_{\beta}}{2\alpha\Psi^4}||.
\end{equation}
Here $R_\Psi$ and $R_\beta$ denote the residuals of Eqs.~(\ref{eq:XCTS-ConformalFactor}) and~(\ref{eq:XCTS-Shift}), respectively, and $||\,.\,||$ represents the root-mean-square value over grid-points of
the
 entire computational grid. 
This plot demonstrates that our initial
data solver converges  exponentially with resolution, even for
very high spins, which gives confidence that we are indeed correctly
solving the Einstein Field Equations.

\begin{figure}
\includegraphics[width=0.95\columnwidth]{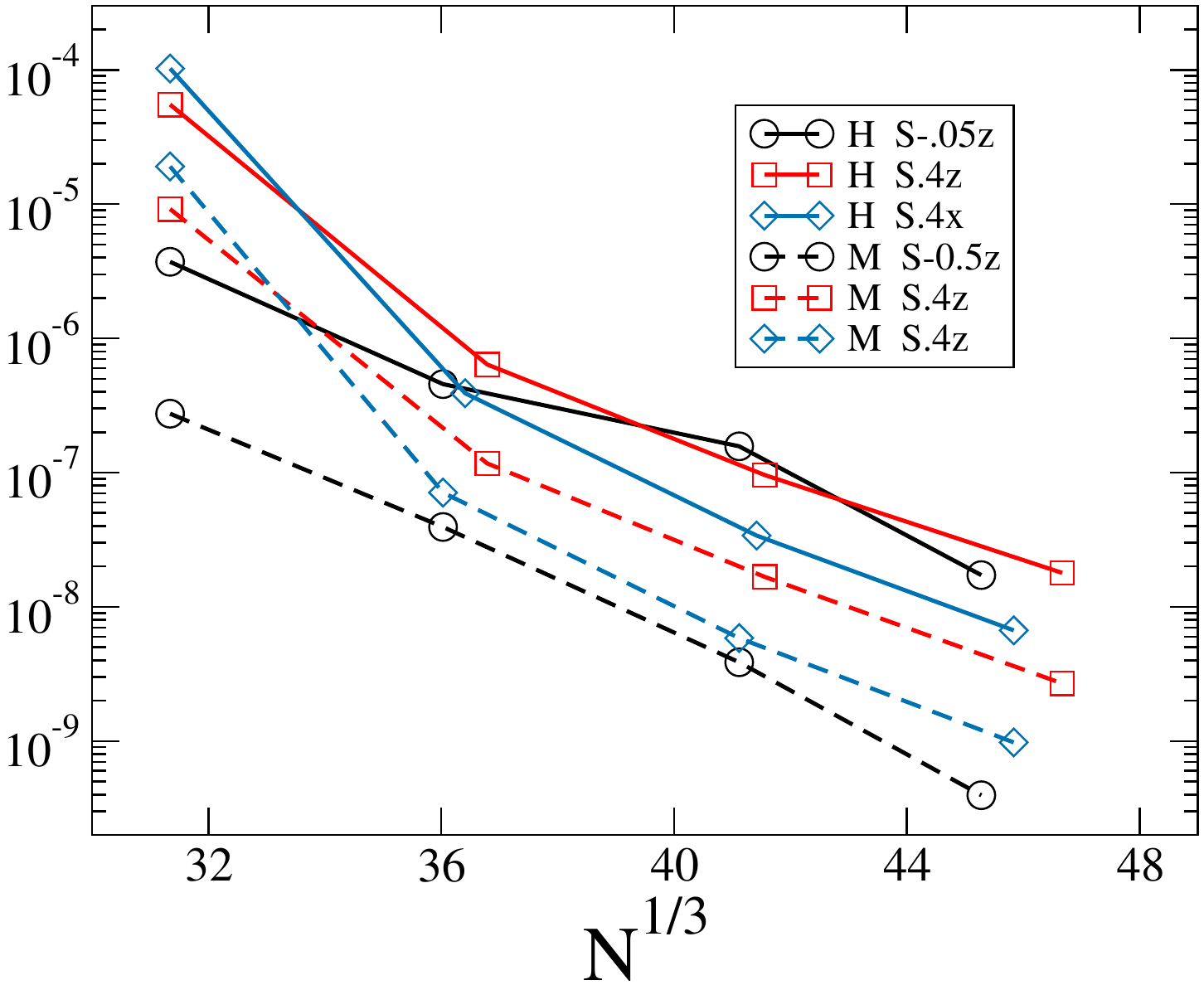}
\caption{{\label{fig:HamMom}}Hamiltonian and Momentum constraints as a function of resolution $N$. We see exponential
  convergence in all cases.}
\end{figure}

 The surface of the star is represented by a spherical harmonic expansion:
\begin{equation}
R_s(\theta,\phi) = \sum^{l_{\rm max},m_{\rm
    max}}_{l,m}c_{lm}Y_{lm}(\theta,\phi),
\end{equation}
where $l_{\rm max}= m_{\rm max} = 11$, unless stated otherwise.
The stellar surface is located by finding a constant enthalpy surface, cf. Sec.~\ref{sec:IDalgorithm}, and the spherical subdomains that cover the star are deformed to conform to $R_s(\theta,\phi)$.
To establish convergence of the position of the stellar surface we
introduce the quantity
\begin{equation}
\label{eq:Deltac}
\Delta c(i) = \frac{1}{l(l+1)} \sqrt{\sum^{l_{\rm max},m_{\rm
      max}}_{l,m}\left(c_{lm}(i)-c_{lm}(N)\right)^2}.
\end{equation}
Here $i$ refers to the $i^{\rm th}$ resolution in the initial data,
and $N$ refers to the final resolution. Figure~\ref{fig:ClmDif} plots $\Delta c(i)$ vs. resolution.  The surface location converges exponentially to better than $10^{-8}$.

\begin{figure}
\includegraphics[width=0.95\columnwidth]{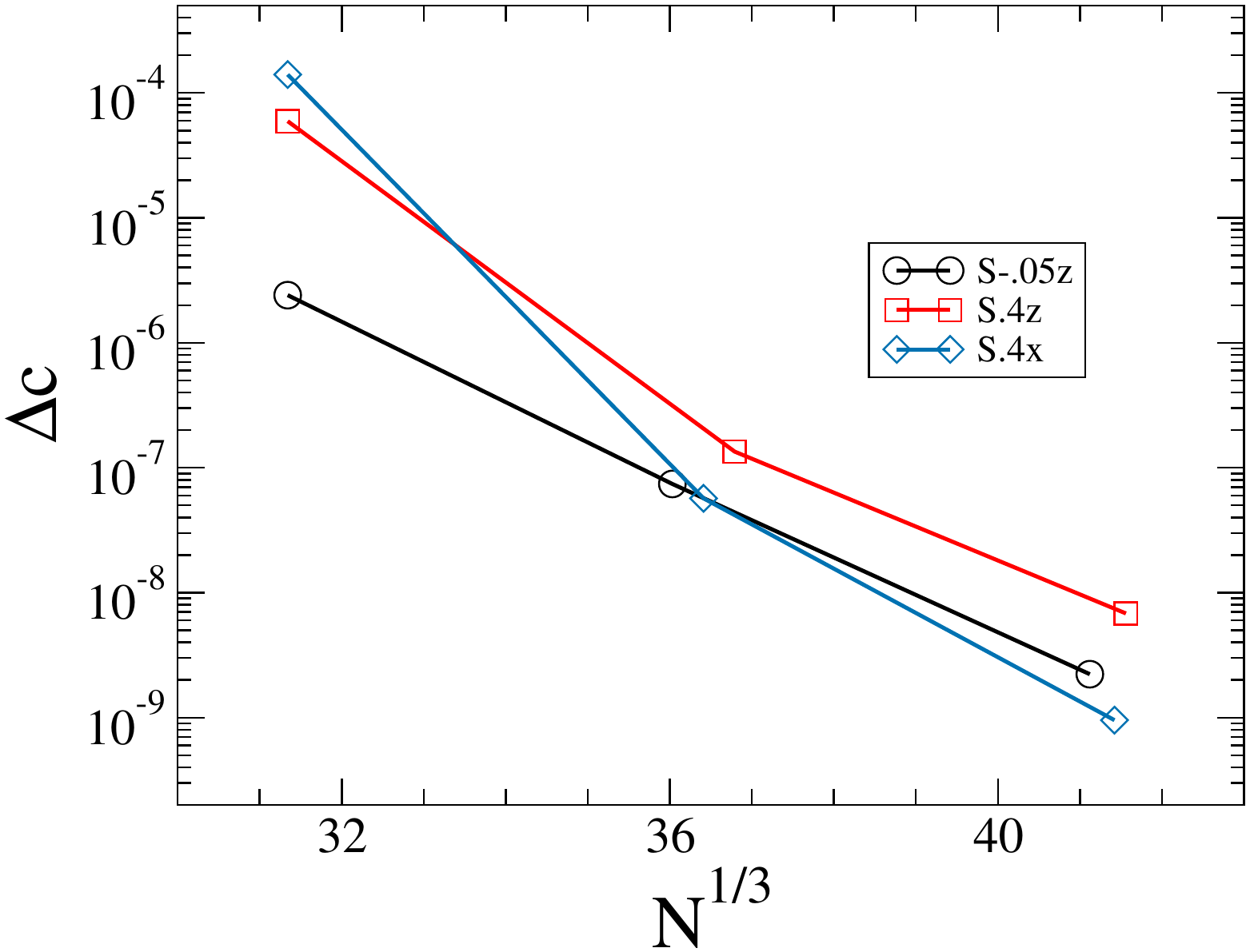}
\caption{{\label{fig:ClmDif} Convergence of the location of the stellar surface}.
  Plotted is $\Delta c$ as defined in Eq.(~\ref{eq:Deltac}), for three representative configurations.}
\end{figure}

Finally, we assess the overall convergence of the solution through
the global quantities $E_{\rm ADM}$ and $\left|J^i_{\rm ADM}\right|$.
The surface integrals at infinity in these two quantities are recast using Gauss' law (cf.~\cite{FoucartEtAl:2008}):
\begin{align}
E_{\rm ADM} &= -\frac{1}{2\pi}\oint_{S_{\infty}}\delta^i_j\partial_i\Psi\, dS_j\nonumber \\
&= -\frac{1}{2\pi}\oint_{S}\delta^i_j \partial_i\Psi\, dS^j  +\frac{1}{2\pi}\int_{\mathcal{V}}\delta^{ij}\partial_i\partial_j\Psi\,dV,
\end{align}
and
\begin{align}
J^{z}_{\rm ADM} 
& = \frac{1}{8\pi}\oint_{S_{\infty}}\left(xK^{yj}-y K^{xj}\right)dS_j\nonumber\\
& = \frac{1}{8\pi}\oint_{S}\left(xK_{yi}-yK_{xi}\right)\delta^{ij}\Psi^2\,dS_j.
\end{align}
Here $\mathcal V$ is the volume outside $S$, and the integrals are evaluated in the flat conformal space.  To obtain the other components of $J_{\rm ADM}^i$, cyclically permute the indices x,y,z.
 We define the quantities $\Delta E$ and $\Delta J$ as the absolute fractional difference in these quantities between the current resolution and the next highest resolution. These are plotted in figure~\ref{fig:EADMConvergence}.   
In general, we find
agreement at the $10^{-7}-10^{-8}$ level by the final
resolution.

\begin{figure}
\includegraphics[width=0.95\columnwidth]{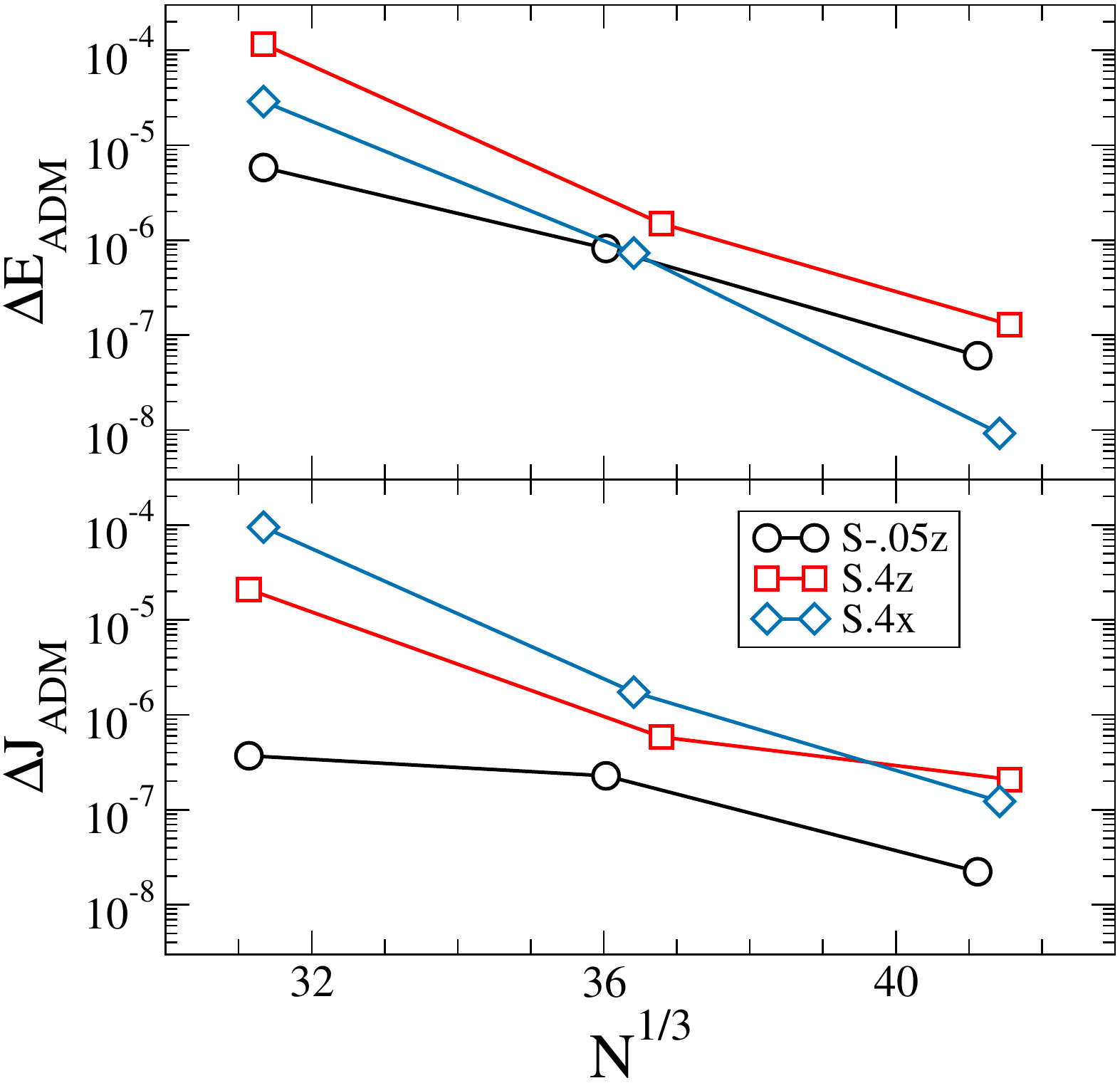}
\caption{{\label{fig:EADMConvergence}} Convergence of ADM-energy and the magnitude of the ADM-angular momentum.  Shown are the fractional differences between neighboring resolutions, as a function of the lower resolution.
}
\end{figure}

\subsection{Convergence of the quasi-local spin}

\begin{figure}
\includegraphics[width=0.95\columnwidth]{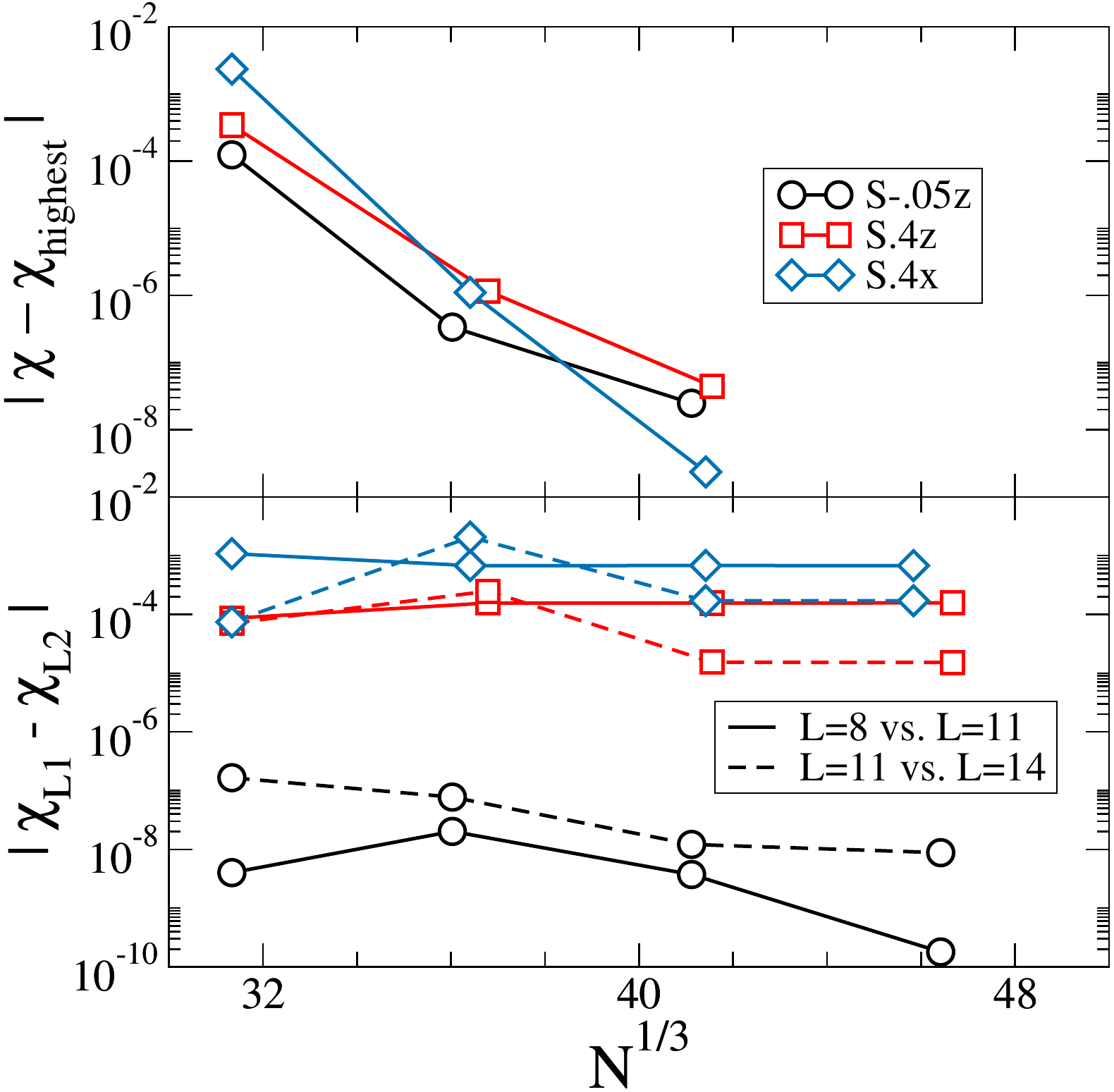}\\

\caption{\label{fig:SpinConvergence} Convergence of the quasi-local
  spin computation.  {\bf Top panel:} difference of spin computed at
  resolution $N$ with the spin computed at the highest resolution.
  {\bf Bottom panel:} Difference between spins computed at different
  resolution $L$ of the spin-computation.  For {\tt S-.5z}, we achieve an
  accuracy of $\sim 10^{-7}$, whereas for {\tt S.4z} and {\tt S.4x}, the accuracy
  is $\sim 10^{-4}$ due to finite $L$.}
\end{figure}

We now turn to the angular momentum of the neutron stars, as measured
with quasi-local angular momentum integrals on the stellar surface.
We will discuss dimensionless spins $\chi$, which depend on two
distinct numerical resolutions: First, the resolution of the
3-dimensional grid used for solving the initial value equations.  This
resolution is specified in terms of $N$, the total number of
grid-points.  Second, the resolution used when solving the eigenvalue
problem for approximate Killing vectors on the 2-dimensional surface,
as given by $L$, the expansion order in spherical harmonics of
  the surface-parameterization
  $r_S(\theta,\phi)=\sum_{l=0}^L\sum_m r_{lm}Y^{lm}(\theta,\phi)$.

Throughout this paper, we use $L=11$.  The top panel of
Fig.~\ref{fig:SpinConvergence} shows convergence of $\chi$ with
grid-resolution $N$, at fixed $L=11$.  We find near exponential
convergence.

The influence of our choice $L=11$ is examined in the lower panel by
computing the quasi-local spin at lower resolution $L=8$ and at higher
resolution $L=14$.  Changing $L$ impacts $\chi$ by $\sim 10^{-8}$ for
the low-spin case {\tt S-.05z}, and by $\sim 10^{-4}$ for the high-spin
cases {\tt S.4z} and {\tt S.4x}.  For the high-spin cases, the spin
  measurement is convergent with increasing $L$, and the finite value
  of $L$ dominates the error budget.  For the low-spin case, numerical
  truncation error dominates the error budget and convergence with $L$
  is not visible.  High NS spin leads to a more distorted stellar
  surface, and so a fixed $L=11$ yields a spin result of lower
  accuracy.  However, in all cases the the numerical errors of our
  spin measurements are still neglible for our purposes.

\begin{figure}
\includegraphics[width=0.98\columnwidth]{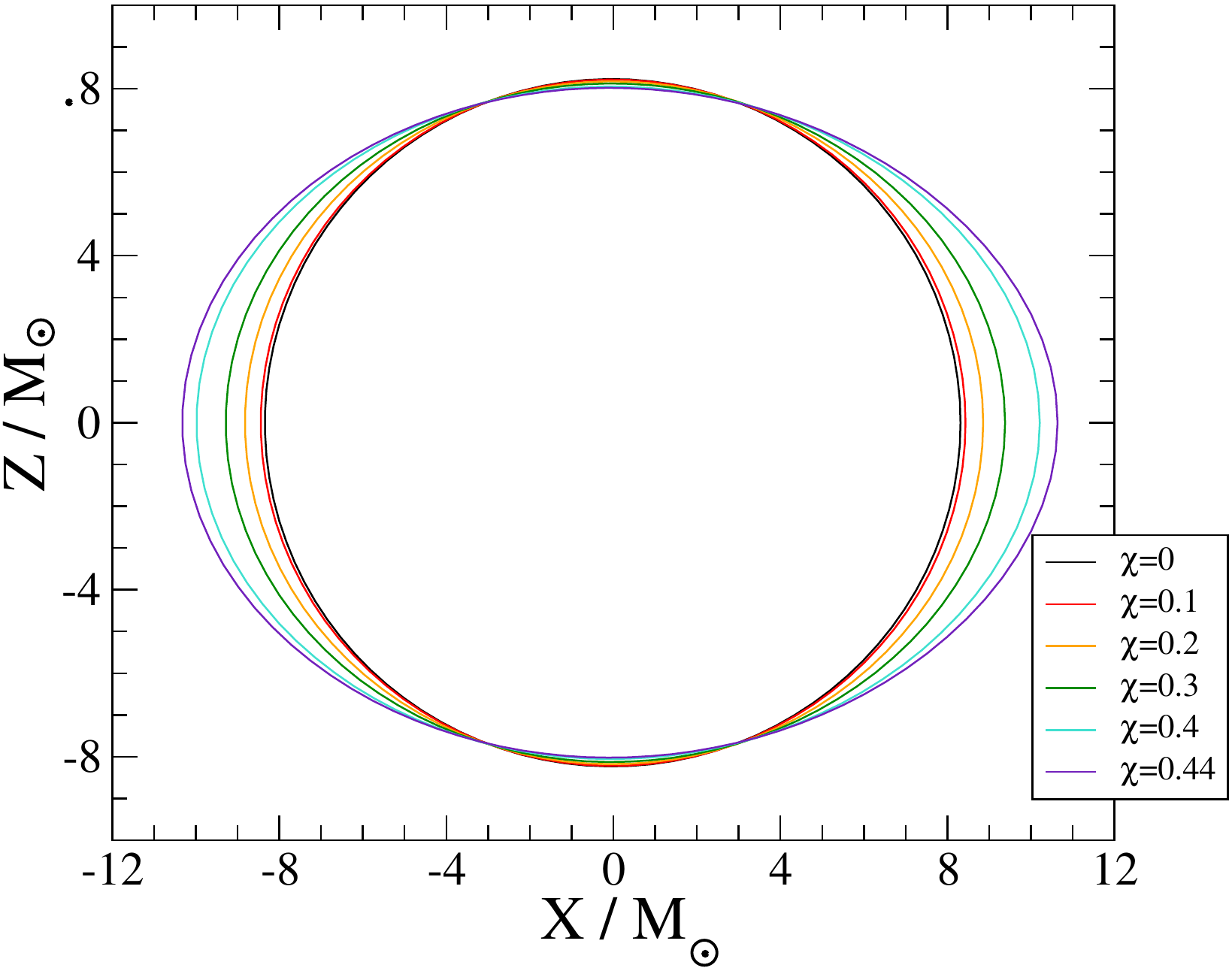}
\caption{{\label{fig:Bulging}}Stellar cross-sections in the X-Z plane for a
  series of different spins, aligned with the $\hat{z}$ axis,
  demonstrating that they bulge at the equator in the expected way
  with increasing spin.}
\end{figure}

\subsection{Quasi-local Spin}
\label{sec:QLSpinProperties}

As discussed in section~\ref{sec:QLSpinExplanation}, we use a
quasi-local spin to define the angular momentum carried by each
neutron star.  To our knowledge, this is the first application of this
method to neutron stars in binaries. 

In this section, we explore properties of the rotating BNS initial
datasets and the employed quasi-local spin diagnostic.

To explore the spin-dependence of BNS initial data sets, we
construct a sequence of equal-mass, equal-spin BNS binaries, with
spins parallel to the orbital angular momentum.  We fix the initial
data parameters $M^b_{\rm NS}$, $D_0$, $\Omega_0$ and $\dot{a}_0$ to
their values for a configuration that we will also evolve below (specifically,  {\tt S.4z} - Ecc1)

\begin{figure}
\includegraphics[width=0.95\columnwidth]{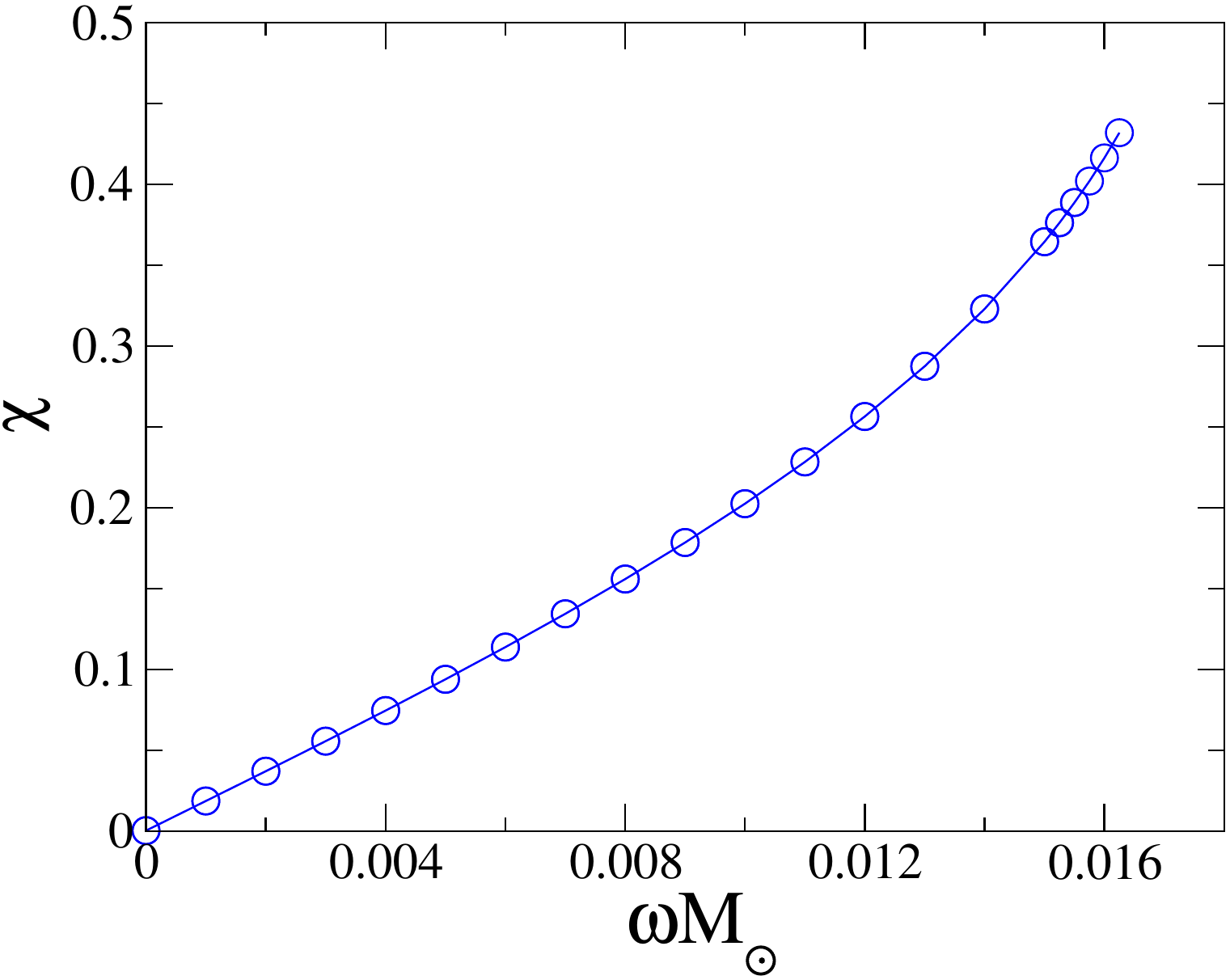}
\caption{{\label{fig:ChiVOmega}}Dimensionless angular momentum $\chi$ as a function of $\Omega$ for
  a series of spin-aligned initial data sets with the same physical
  parameters as our runs of interest. We see, as expected, a linear
  relation between $\chi$ and $\Omega$ at low-spins, which eventually
  becomes non-linear at higher spins. }
\end{figure}

Figure~\ref{fig:Bulging} shows cross-sections through one of the neutron stars
in the xz-plane, i.e. a plane orthogonal to the orbital plane which is
intersecting the centers of both stars.  With increasing spin, the
stars develop an increasing equatorial bulge, an expected consequence of centrifugal forces. 

Figure~\ref{fig:ChiVOmega}
presents the dimensionless spin of either neutron star as a function
of $\omega$.  $\chi$ increases monotonically with the rotation
parameter $\omega$.  
The spin $\chi$ increases
linearly with $\omega$ for small $\omega$.  For larger $\omega$, the
dependence steepens, as the increasing equatorial radius of the stars
increase the moment of inertia~\cite{Worley:2008cb}. 

For
$\omega=0.01625 M_{\odot}^{-1}$ we achieve $\chi=0.432$, the largest spin we are
able to construct.  This is reasonably close to the theoretical
maximum value for $\Gamma=2$ polytropes,
$\chi\sim 0.57$~\cite{Ansorg:2003br}.  Above $\omega=0.01625 M_{\odot}^{-1}$,
the initial data code fails to converge.  The steepening of the $\chi$
vs. $\omega$ curve is reminiscent of features related to
non-uniqueness of solutions of the extended conformal thin sandwich
equations~\cite{Lovelace2008,Pfeiffer-York:2005,Baumgarte2007,Walsh2007},
and it is possible that our failure to find solutions originates in an analogous
break-down of the uniqueness of solutions of the constraint equations.

While the focus of our investigation lies on rotating NS, we
  note that for $\omega=0$ our data-sets reduce to the standard
  formalism for irrotational NS.  For $\omega=0$, we find a
  quasi-local spin of the neutron stars is $\chi=2\times 10^{-4}$.
  This is the first rigorous measurement of the residual spin of
  irrotational BNS.  Residual spin is, for instance, important for the
  construction and validation of waveform models for compact object
  binaries.  The analysis in Ref.~\cite{Boyle2007} indicates that
  spins of order $10^{-4}$ lead to a dephasing of about 0.01radians
  during the last dozen of inspiral orbits.  This value is
  significantly smaller than the phase accuracy obtained by current
  BNS simulations, and so the residual spin is presently not a
  limiting factor for studies like~\cite{Bernuzzi:2014owa,Baiotti2011,Baiotti:2010xh}.

Finally, we demonstrate that the surface on which we
compute the quasi-local spin, does not significantly impact the spin
we measure: We choose coordinate spheres centered on the neutron star
with radius $R$, and compute the quasi-local spin using these
surfaces,
rather than the stellar surface.

In Fig.~\ref{fig:ChiVR}, we plot the spin measured on various $R=\rm{const}$
surfaces, for three different values of $\omega$, from the same sequences
shown in Fig.~\ref{fig:ChiVOmega}.

The circles denote spins extracted on coordinate spheres.  The asterisks
indicate the spins computed on the stellar surface. The asterisk is
plotted at $R=R_{\rm eq}$, the equatorial radius of the neutron star
under consideration.  
We find good agreement between spins extracted on coordinate spheres and
the spin extracted on the stellar surface, as long as $R\ge R_{\rm eq}$. 
 The maximum disagreement is seen in the high spin curve, where
the two spins differ by $\sim 10^{-2}$.  

For $R<R_{\rm eq}$, the coordinate extraction sphere intersects the
outer layers of the neutron star and no longer encompasses the entire matter and angular
momentum of the star.  Therefore, $\chi(R)$ shows a pronounced decline
for $R<R_{\rm eq}$ for each of the three initial-data sets considered
in Fig.~\ref{fig:ChiVR}.  For $R>R_{\rm eq}$, $\chi(R)$ continues to
increase slightly, for instance, for the middle curve,
$\chi(R=9)=0.202$ whereas $\chi(R=11)=0.204$.

In summary, Fig.~\ref{fig:ChiVR} shows that the quasi-local spin
extracted on coordinate spheres can serve as a good approximation of
the quasi-local spin extracted on the stellar surface (as long as the
coordiate sphere is outside the star, of course).

This is important because during evolutions of the
binary, we do not track the surface of the star.
Instead, we will compute the spin on coordinate spheres, similarly to Fig.~\ref{fig:ChiVR}.

\begin{figure}
\includegraphics[width=0.95\columnwidth]{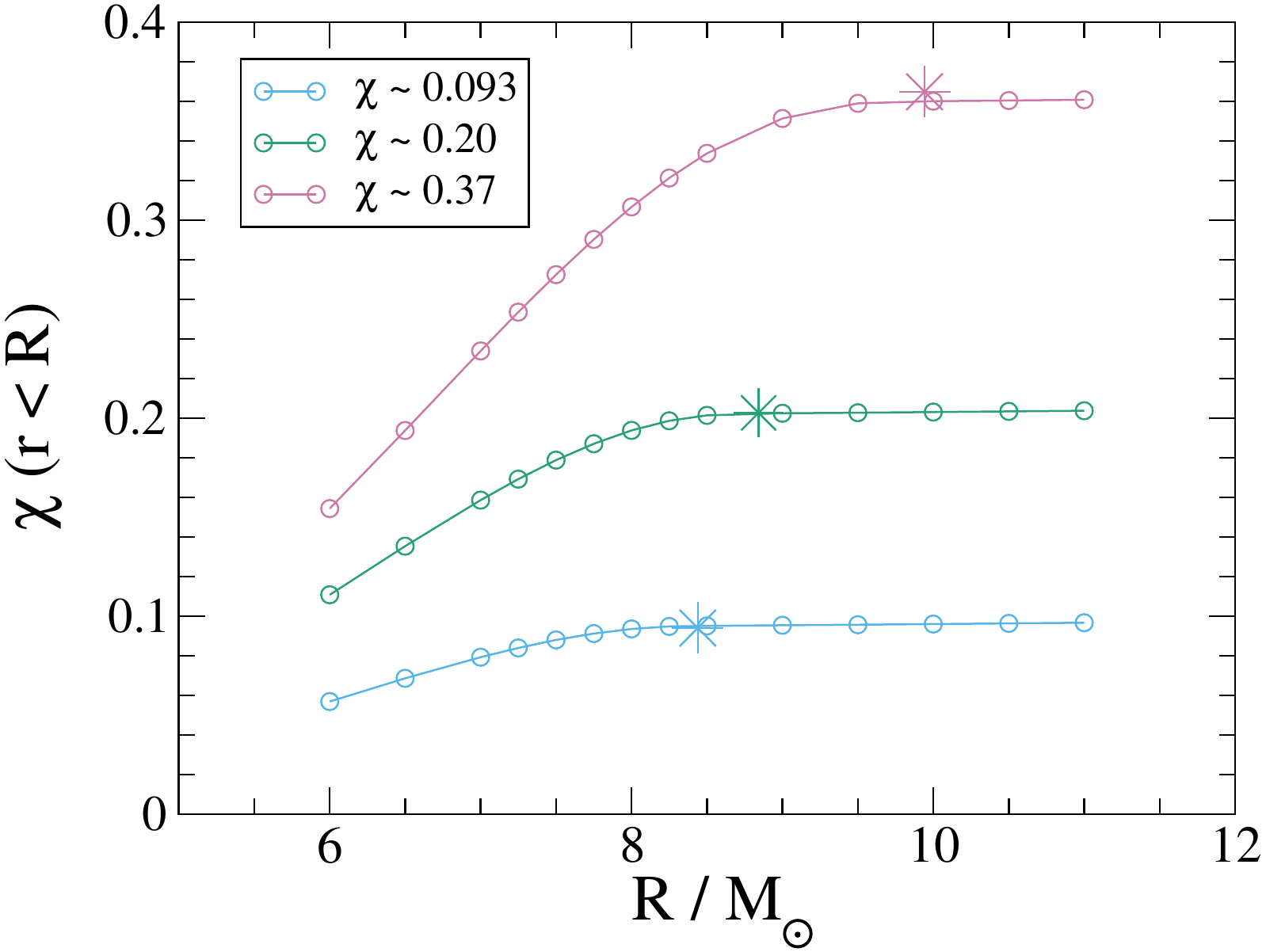}
\caption{{\label{fig:ChiVR}}Dimensionless spin $\chi$ measured on
  coordinate spheres with radius $R$ for three different aligned spin
  BNS systems.  The asterik denotes the spin measured on the
  (non-spherical) stellar surface.  Circles to the right of the
  asterik represent coordinate spheres entirely outside the neutron
  star, and circles on the left of the asterik indicate spin
  measurement surfaces that intersect the star or are entirely located
  inside the star.}
\end{figure}

\section{Evolution Results}
\label{sec:EvolutionResults}

We now evolve the three configurations discussed in Sec.~\ref{sec:ID}.
As indicated in Table~\ref{tab:InitialData}, all three configurations
are equal-mass binaries, with individual ADM masses $M_\star$ (in
isolation) 
of $1.64M_{\odot}$ or $1.648M_{\odot}$ at initial separation of
$D=47.2M_\odot$, and using a polytropic equation of state with
$\Gamma=2.0$ and $\kappa=123.6$.  Both stars have equal spins, and the
three configurations differ in spin magnitude and spin direction.
Configuration {\tt S-0.05z} has spin-magnitudes $\sim 0.05$ anti-aligned
with the orbital angular momentum, and the confiurations {\tt S.4z} and {\tt S.4x}
have spin magnitudes near 0.4, along the z-axis and x-axis,
respectively.

Each configuration is evolved through $\gtrsim 10$ orbits, into the
late-inspiral.  In this paper we focus on the inspiral of the neutron
stars.  Table~\ref{tab:RunInfo} summarizes parameters for these runs.

\begin{table}
\begin{tabular} {c | c | c | c | c | c | c}
Name & $k$ & $e$ & $\vec{\chi}$ & $f_{0}(Hz)$ & $N_{\rm orb}$ & $t_{f}({\rm ms})$ \\ \hline
{\tt S.4z} & 0,1,2 & $\lesssim 0.001$ & $0.381\hat{z}$ & $167.7$ & $11.8$ & $56.0$ \\ \hline
{\tt S-.05z} & 0,1,2 & $0.0006$ & $-0.050\hat{z}$ & $165.4$ & $12.5$ & $56.3$ \\ \hline
{\tt S.4x} & 0,1 & $\lesssim 0.002$ & $0.375\hat{x}$ & $164.8$ & $9.1$ & $45.7$ \\ \hline
\end{tabular}
\caption{ {\label{tab:RunInfo}} Information about our three evolutions.  $k$ indicates the numerical resolutions on which a simulation is performed, $e$ indicates the smallest achieved orbital eccentricity.  $\vec\chi$ and $f_0$ are the dimensionless spins at $t=0$ and the initial orbital frequency.  Finally, $N_{\rm orb}$ and $t_f$ represent the number of orbits the configuration was evolved for, and the evolution time.}
\end{table}

\subsection{Evolution Code}
\label{sec:EvolutionCode}

In our evolution code, SpEC~\cite{Buchman:2012dw,Lovelace:2011nu,
  Scheel2009,Kidder2000a,Lindblom2006,Scheel2006,Szilagyi:2009qz,Lovelace:2010ne,
  Hemberger:2012jz,Ossokine:2013zga}, we use a mixed spectral --
finite-difference approach to solving the Einstein Field Equations
coupled to general relativistic hydrodynamics equations.  The
equations for the space-time metric, $g_{\mu\nu}$ are solved on a
spectral grid, while the fluid equations are solved on a finite
difference grid, using a high-resolution shock-capturing scheme. We
use a WENO~\cite{Jiang1996202,Liu1994200} reconstruction method to
reconstruct primitive variables, and an HLL Riemann solver~\cite{HLL}
to compute numerical fluxes at interfaces. Integration is done using a
3rd order Runge-Kutta method with an adaptive stepsize. We interpolate
between the hydro and spectral grids at the end of each full time
step, interpolating in time to provide data during the Runge-Kutta
substeps
(see~\cite{Duez:2008rb,FoucartEtAl:2011,Foucart:2013a,Muhlberger2014}
for a more detailed description of the method).  

Each star is contained in a separate cubical finite
difference grid that does not overlap with that
of the other star.  The sides of the grids are initially $1.25$ times
the stars' diameters. We use grids that contain $97^3$, $123^3$ and
$155^3$ points for resolutions $k=0,1,2$, respectively\footnote{For
  aligned-spin configurations {\tt S-.05z} and {\tt S.4z}, we take advantage of,
  and enforce, z-symmetry, which halves the number of grid-points
  along the z-axis.}.  These resolutions correspond to linear
grid-spacing of $340\,\text{m}$, $268\,\text{m}$ and $213\,\text{m}$
respectively for the {\tt S.4z} case.  The precessing evolution {\tt S.4x} uses
similar grid-spacing, whereas the anti-aligned run {\tt S-.05z} has a
slightly smaller grid-spacing because the stars themselves are
smaller. The region outside the NS but inside the finite
difference grid is filled with a low density
atmosphere with $\rho=10^{-13}M_{\odot}^{-2}$.  The motion of the NSs
is monitored by computing the centroids of the NS mass distributions
\begin{equation}
\label{eq:centroid}
X^i_{\rm CM} = \int{x^iu^0\rho_0\sqrt{-g^{(4)}}d^3x}
\end{equation}
for each of the grid patches containing a NS.

The grids are rotated and their separation rescaled to keep the
centers of the NS at constant grid-coordinates~\cite{Scheel2006,Hemberger:2012jz,Scheel2014}.  As the physical separation between the stars decreases, the rescaling of
grid-coordinates therefore causes the size of the stars to increase
in grid-coordinates.  In order to avoid the stellar surfaces expanding
beyond the geometric size of the finite difference grid, we monitor the matter flux leaving
this grid along the x, y, and z-direction.  If the matter flux is too
large along a certain axis, we expand the grid in that
direction.  
This procedure allows us to
dynamically choose the optimal grid-size that limits matter loss to a
small, user-specified level.  When changing the size of the hydro
grid, the number of grid-points is kept constant, so this process
changes the effective resolution during the evolution.

The Einstein field equations are solved on a spectral grid using basis-functions appropriate for the shape of each subdomain.
For rectangular blocks, Cheybyshev polynomials are used along each axis; for a spherical shell (i.e. where the center is excised), spherical harmonics in angles, and Chebyshev polynomial in radius are employed; and for an open cylinder (i.e. with the region near the axis excised), Chebyshev polynomials and a Fourier series.  For full spheres and filled cylinders, multi-dimensional basis-functions respecting the continuity conditions at the orign/axis are employed~\cite{Matsushima-Marcus:1995,1997JCoPh.136..100V}.  For more details see~\cite{Muhlberger2014}.

More specifically, our spectral
grid, the central region of each star is covered by a filled sphere
located at the center of the star. These have spherical harmonic modes
up to $L = 12+2k$.  The radial basis-functions are one-sided Jacobi
polynomials with $7+k$ collocation points. The filled spheres are
surrounded by eight other spherical shells with the same radial and
angular resolutions. At the start of the evolution, the stellar
surface is generally located inside the third shell.  The far field
region is covered by 20 spherical shells starting at 1.5 times the
inital binary separation and going out to 40 times that
separation. These shells have angular resolution $L=9+2k$ and radial
resolution $6+k$.  The region between the innermost shell and the
stars is covered by a set of cylindrical shells and filled cylinders.

We use a generalized harmonic evolution system~\cite{Pretorius2006,Pretorius2005a,Lindblom2006} with 
 coordinates $x^{\mu}$ such that they satisfy a
wave equation
\begin{equation}
\nabla^{\nu}\nabla_{\nu}x^{\mu} = H^{\mu},
\end{equation}
for some freely-specifiable source function $H^{\mu}$. The initial
source function $H^\mu_{\rm initial}$ is determined by the intial
data, assuming that the time derivatives of the lapse and
shift functions initially vanish in the corotating frame. 
We then transition to a pure harmonic gauge, $H^{\mu}=0$ by
using a transition function, i.e.
\begin{equation}
H^\mu = e^{-\left(t/\tau\right)^4}\; H^\mu_{\rm initial}.
\end{equation}
The timescale $\tau$ is determined by $\tau=2\sqrt{d^3/(2M_\star)}$.
This is slow enough to avoid numerical gauge artifacts in the simulations.

\subsection{Eccentricity Removal}
\label{sec:EccRemoval}

Gravitational wave emission reduces orbital eccentricity rapidly during
a GW-driven inspiral~\cite{PetersMathews1963,Peters1964}.  Therefore
inspiraling binary neutron stars are expected to have essentially
vanishing orbital eccentricity in their late inspiral, unless they recently
underwent dynamical interactions.  Our goal is to model
non-eccentric inspirals.  In this subsection we demonstrate that we can
indeed control and reduce orbital eccentricty, using the techniques
developed for BH-BH
binaries~\cite{,Pfeiffer-Brown-etal:2007,Boyle2007,Buonanno:2010yk}
and also applied to BH-NS binaries~\cite{FoucartEtAl:2008}.

\begin{figure}
  \includegraphics[width=0.95\columnwidth]{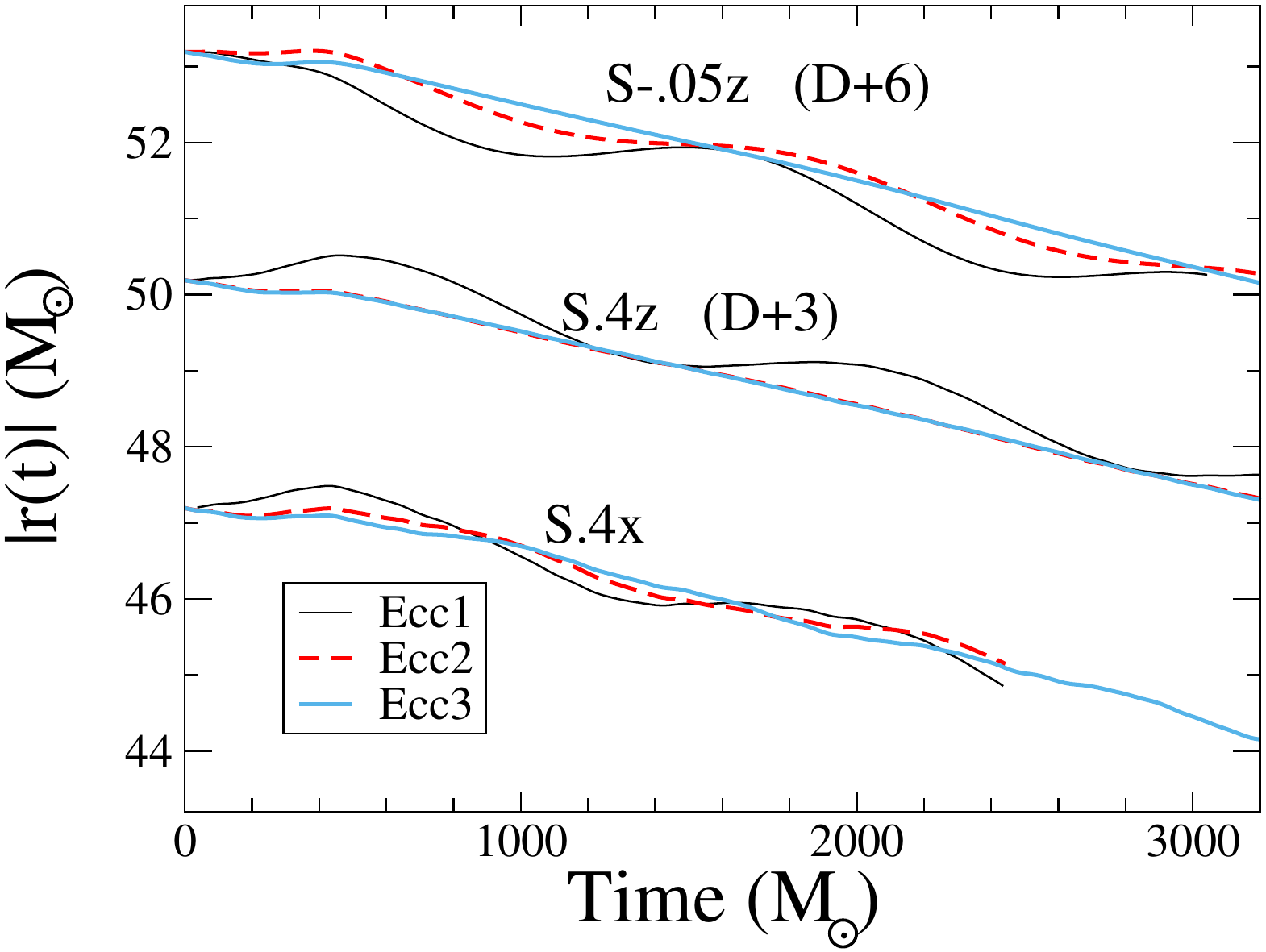}\caption{
{\label{fig:DvT}}
The
    binary separation as a function of time.  Shown are three
    eccentricity removal iterations (Ecc1,Ecc2,Ecc3) for each of the
    three configurations studied.  The data for {\tt S-.05z} and {\tt S.4z} is
    offset vertically by 6 and 3, respectively, for clarity of
    plotting.}
\end{figure}

For fixed binary parameters (masses, spins), and fixed initial
separation $D_0$, the initial orbit of the binary is determined by two
remaining parameters: The initial orbital frequency $\Omega_0$, and
the initial radial velocity, which we describe through an expansion
parameter $\dot{a}_0 = \dot{r}/r$.  These two parameters will encode
orbital eccentricty and phase of periastron, and our goal is to
determine these parameters to reduce orbital eccentricity.  We
accomplish this using an iterative procedure first introduced
  for binary black holes~\cite{Boyle2007,Buonanno:2010yk}. An initial
data set is evolved for a few orbits, the resulting orbital dynamics
are analyzed, and then the initial data parameters $\Omega_0$ and
$\dot a_0$ are adjusted.

\begin{table}
\begin{tabular} {l | l | c | c}
Name & $\Omega\times10^{3}$ & $\dot{a}_0\times10^{5}$ & $e$
\\ \hline {\tt S.4z} - Ecc1 & $5.10538$ & $0$ & $0.006$ \\ {\tt S.4z} - Ecc2 &
$5.09591$ & $-1.60$ & $\lesssim 0.001$ \\ {\tt S.4z} - Ecc3 & $5.09594$ &
$-1.75$ & $\lesssim 0.001$ \\ \hline {\tt S-.05z} - Ecc1 & $5.10538$ & $0$ &
$0.008$ \\ {\tt S-.05z} - Ecc2 & $5.11561$ & $0$ & $0.004$ \\ {\tt S-.05z} - Ecc3
& $5.11769$ & $-1.71$ & $0.0006$ \\\hline {\tt S.4x} - Ecc1 & $5.10538$ &
$0$ & $0.007$ \\ {\tt S.4x} - Ecc2 & $5.10429$ & $-2.27$ & $0.004$ \\ {\tt S.4x} -
Ecc3 & $5.10064$ & $-2.36$ & $\lesssim 0.002$ \\
\end{tabular}
\caption{\label{tab:ecc_removal} Eccentricity removal for the three
  main runs discussed in this paper.  Only initial orbital frequency
  $\Omega_0$ and initial radial expansion factor $\dot a_0$ are
  changed between different EccN iterations. Recall that these
  quantities have units of $M_{\odot}^{-1}$.}
\end{table}

\begin{figure}
\includegraphics[width=0.95\columnwidth]{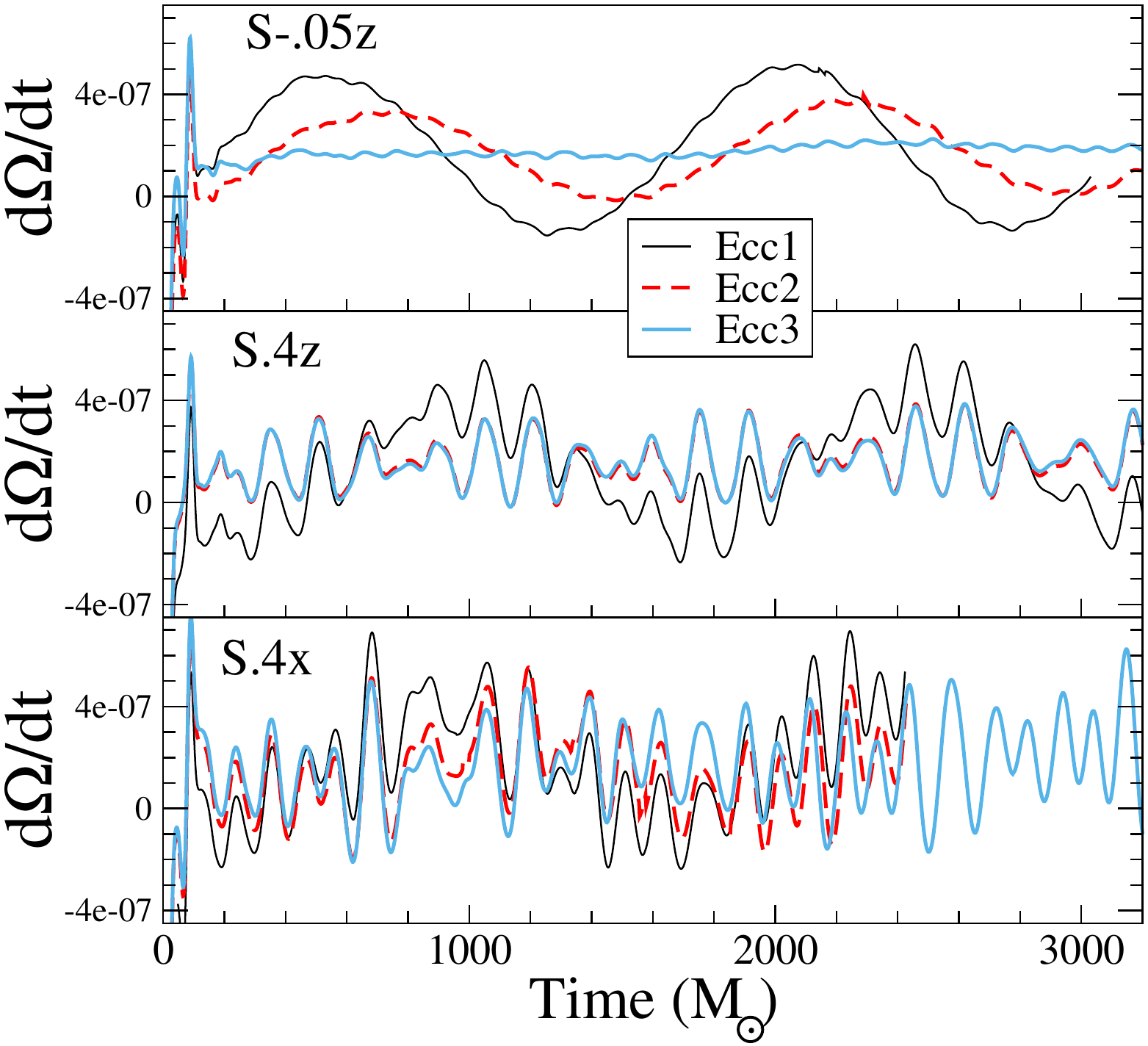}
\caption{{\label{fig:DOmegaVT}} The derivative of the binary orbital
  frequency as a function of time for different levels of
  eccentriccity reduction for our three runs of interest. Note that
  $d\Omega/dt$ has units of $M_{\odot}^{-2}$. }
\end{figure}

For binary neutron stars, we initialize the first iteration
of eccentricity removal, with $\dot{a}_0=0$ and use
$\Omega_0$ determined from irrotational BNS initial data, based on
the equilibrium condition in Eq.~\ref{eq:OmegaEq}.  Evolutions with
these choices are labeled with the suffix ``Ecc1'', and show noticable
variations in the separation between the two NS, cf. the solid black
lines in Fig.~\ref{fig:DvT}.

We
compute the trajectories of the centers of mass of each star, as
  determined by Eq.~\ref{eq:centroid},
$\vec c_1(t)$ and $\vec c_2(t)$, and using the relative separation
$\vec r=\vec c_2(t) - \vec c_1(t)$, compute the orbital frequency
\begin{equation}
\label{eq:Omega}
\Omega(t) = \frac{|\vec r(t)\times\dot{\vec{r}}(t)|}{r(t)^2},
\end{equation}
where an over-dot indicates a numerical time-derivative.
Finally, we compute $\dot{\Omega}(t)$ and fit it to a function of the form
\begin{align}
\dot{\Omega}(t) = & A_1(t_c-t)^{-11/8} + A_2(t_c-t)^{-13/8}\nonumber \\
&+ B_0\cos{(B_1t
  + B_2t^2 + B_3)}.
\end{align}
The power law parts of this fit represent the orbital decay due to the
emission of gravitational waves, while the oscillatory part represents
the eccentric part of the orbit. We then update $\Omega_0$ and
$\dot{a}_0$ with the formulae (see \cite{Buonanno:2010yk} for a
detailed overview)
\begin{align}
  \Omega_0 \leftarrow \Omega_0 - \frac{B_0B_1}{4\Omega_0^2}\sin B_3, \\
  \dot{a}_0\leftarrow \dot{a}_0 +\frac{B_0}{2\Omega_0}\cos B_3.
\end{align}
We repeat this procedure twice, resulting in simulations with suffix
Ecc2 and Ecc3.  Table~\ref{tab:ecc_removal} summarizes the orbital
parameters for the individual simulations, and Figs.~\ref{fig:DvT}
and~\ref{fig:DOmegaVT} illustrate the efficacy of the procedure
through plots of separation and time-derivative of orbital frequency.
The eccentricity is successfully reducued from $e\sim 1\%$ to
$\sim 0.1\%$.  After two eccentricity reduction iterations, variations
in $\dot\Omega(t)$ are so small that they are no longer discernible
from higher-frequency oscillations in $\dot\Omega(t)$,
cf. Fig.~\ref{fig:DOmegaVT}.

\begin{figure}
  \includegraphics[width=0.92\columnwidth]{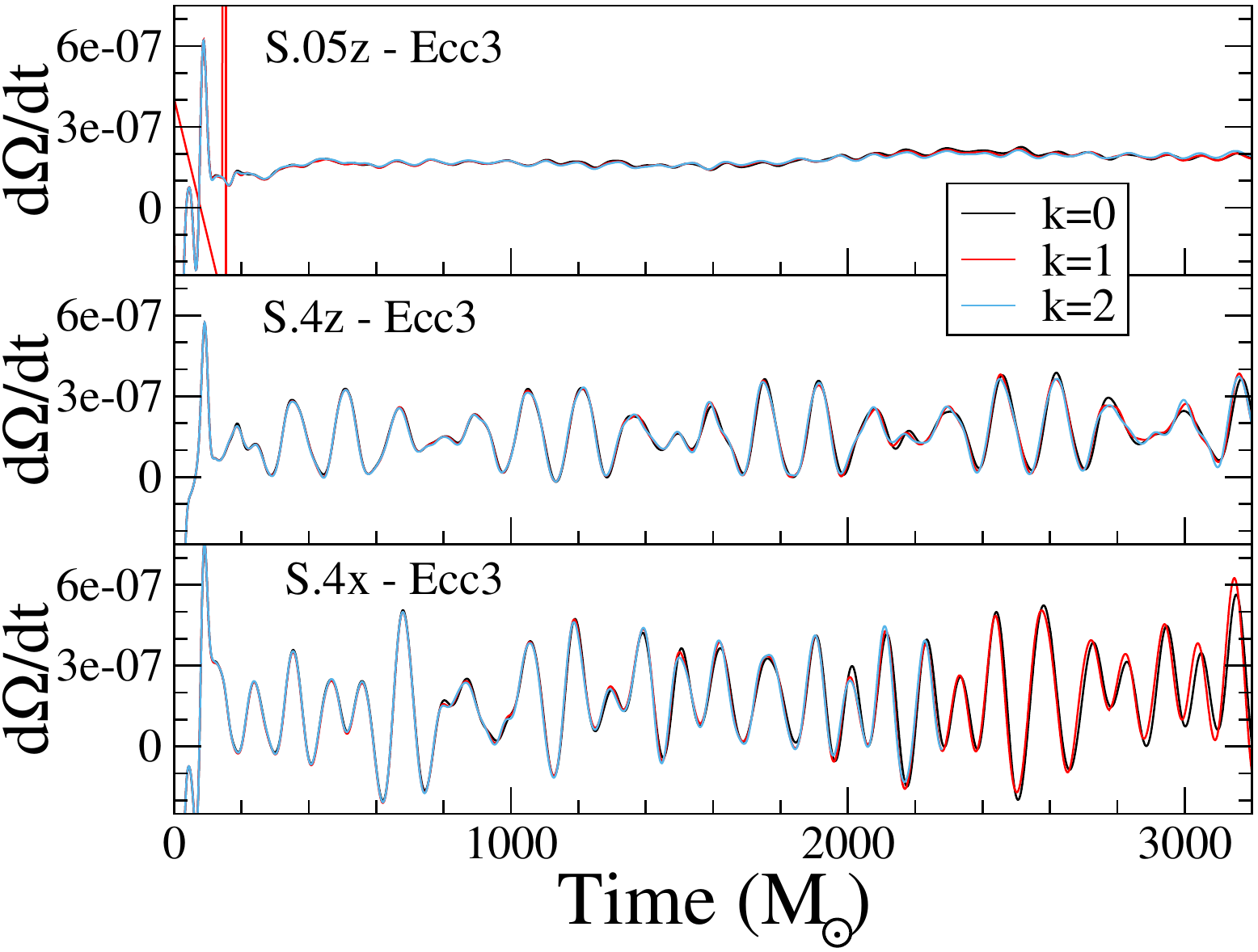}
  \caption{ {\label{fig:OmegaDotComparison}Convergence of
      $\dot\Omega(t)$.  Shown are $\dot\Omega(t)$ at three different
      numerical resolutions ($k=0,1,2$) for the final,
      lowest-eccentricity initial data.  The oscillations in
      $\dot\Omega(t)$ are evidently not caused by numerical truncation
      error. Note that $\dot{Omega}$ has units of $M_{\odot}^{-2}$.}}
\end{figure}

The high freuqency oscillations in $\dot\Omega(t)$ are caused by the
quasi-normal ringing of the neutron stars, as discussed in detail
below in Sec.~\ref{sec:QNModes}.  Here, we only note that these
oscillations are convergently resolved,
cf. Fig.~\ref{fig:OmegaDotComparison}, and are therefore a genuine
feature of our initial data.  Figure~\ref{fig:OmegaDotComparison} also
confirms that the lowest resolution ($k=0$) gives adequate resolution
for eccentricity removal.

The eccentricity removal algorithm attempts to isolate variations on
the orbital time-scale as the signature of eccentricity. For
{\tt S.4z} - Ecc2, it reports $e=0.0005$ and for {\tt S.4z} - Ecc3, $e=0.0002$.
However, given the large amplitude of the QN mode ringing, we consider
these estimates unreliable, and therefore quote an upper bound of
0.001 in Table~\ref{tab:ecc_removal}. Similarly, for {\tt S.4x} - Ecc3, the fitting
reports $e=0.001$, and we quote a conservative upper bound of 0.002.

\subsection{Aligned spin BNS evolutions: NS Spin}
\label{sec:EvolutionSpin}

In this section, we will discuss the measurement of spins during our
evolutions for the non-precessing cases, {\tt S.4z} and {\tt S-.05z}.  Aligned
spin binaries do not precess.  Combined with the low viscosity we
expect the NS spins to stay approximately constant during the
evolutions.  These systems therefore serve as a test on our spin
diagnostics during the evolutions.
In this section, and through the rest of this paper, we always use the
final eccentricity reduction, ``Ecc3''.  For brevity, we will omit the
suffix ``-Ecc3'', and refer to the runs simply as {\tt S-.05z}, etc.

\begin{figure}
\includegraphics[width=0.95\columnwidth]{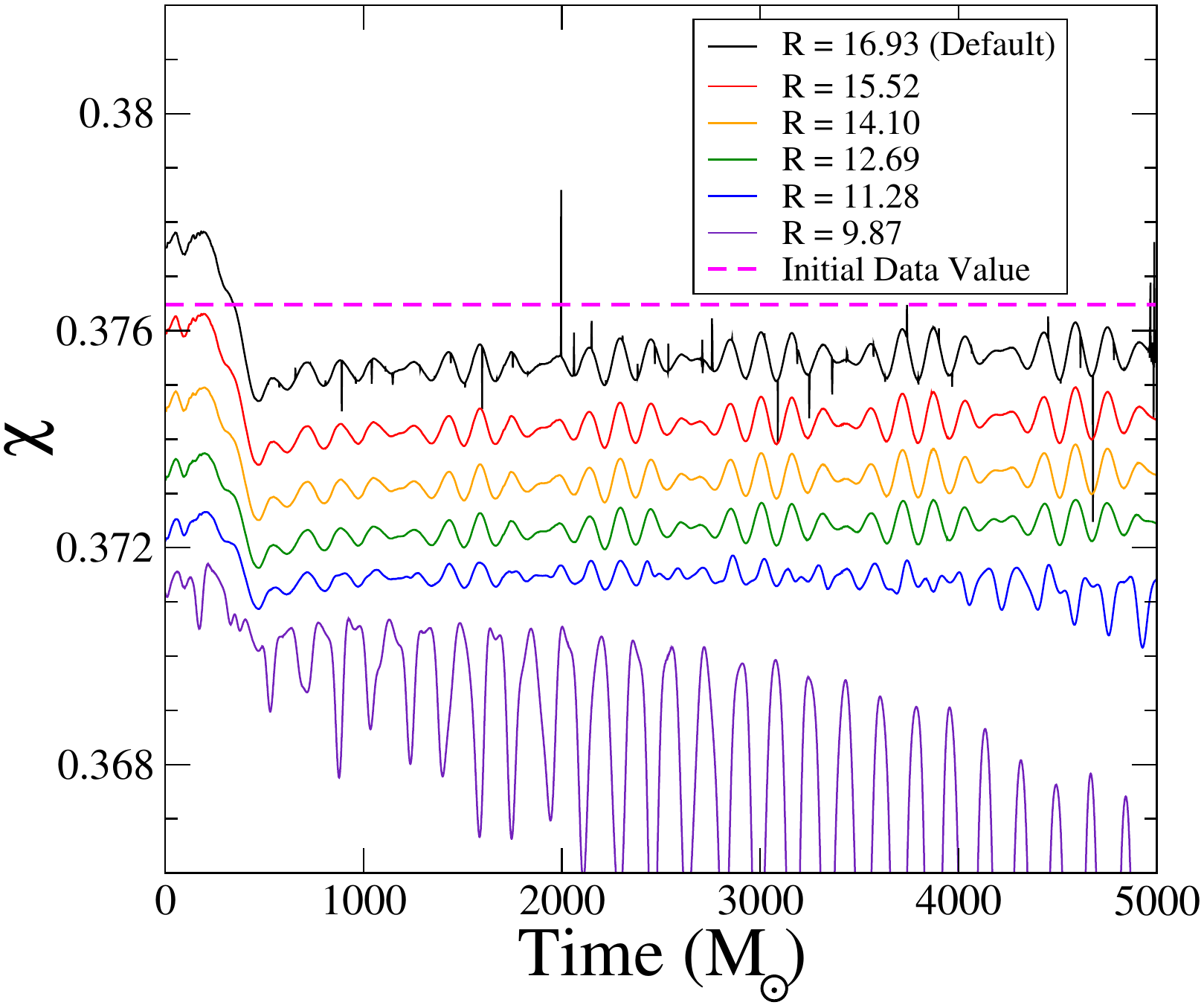}
\caption{{\label{fig:ChiVTZoomed}} The spin measured on multiple
  coordinate spheres for the {\tt S.4z} run.}
\end{figure}

We do not track the surface of the star during the evolution.
Instead we simply evaluate the quasi-local spin of the stars on
coordinate spheres in the frame comoving with the binary.  We must
therefore verify that the spin measured is largely independent of the
radius of the sphere, and that it is maintained during the evolutions
at the value consistent with that in the initial data.
Figure~\ref{fig:ChiVR} established that coordinate spheres can be used
to extract the quasi-local spin in the initial data.
Figure~\ref{fig:ChiVTZoomed} shows the results for the high-spin
simulation {\tt S.4x} during the inspiral.

\begin{figure}
\includegraphics[width=0.99\columnwidth]{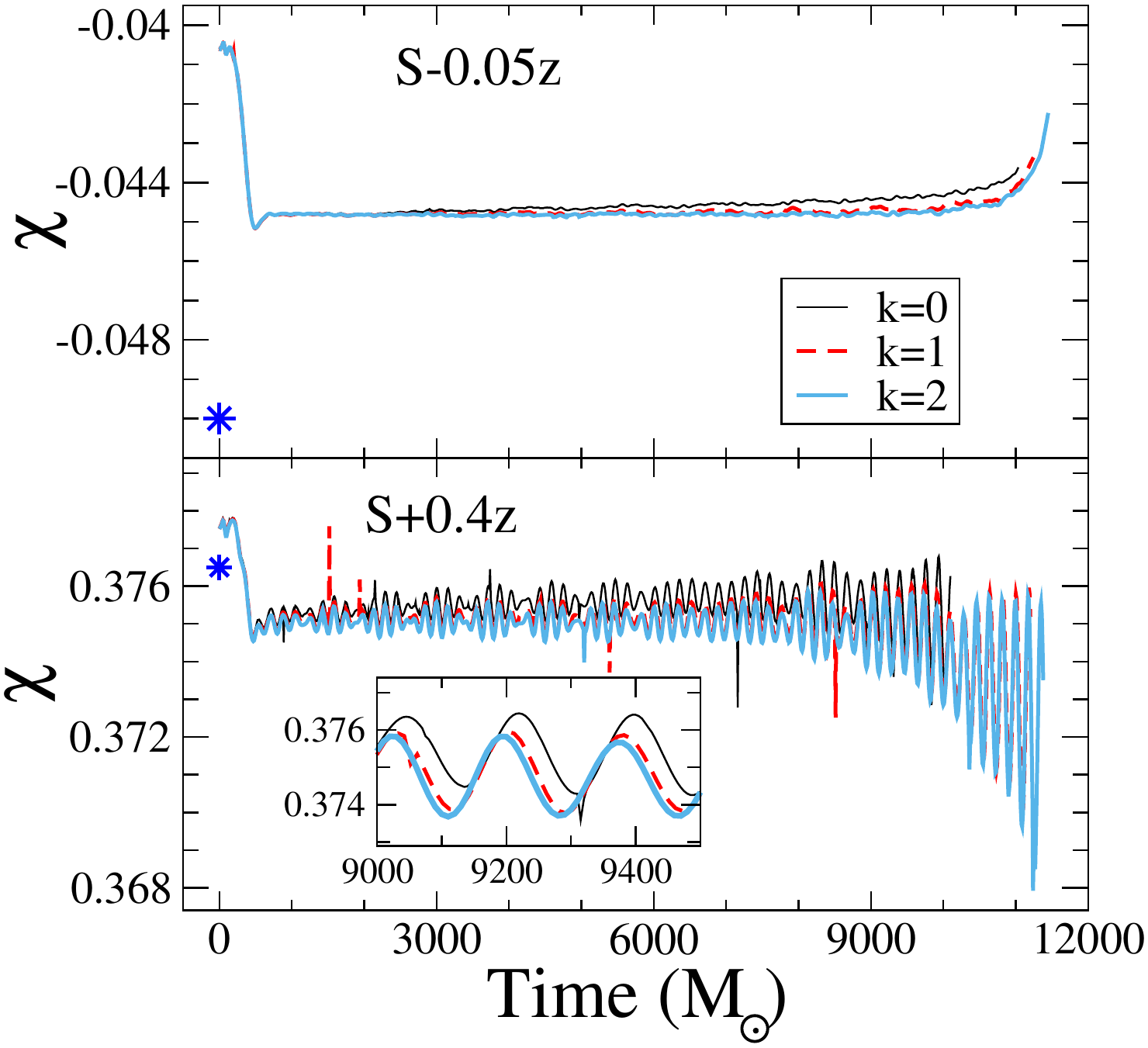}
\caption{{\label{fig:ChiVTDifferentRes2}} Neutron star spin during the
  two aligned-spin evolutions.  Shown are three different numerical
  resolutions, $k=0$ (lowest), $k=1$, and $k=2$ (highest). The
  asterisk indicates the spin measured on the stellar surface in the
  initial data.}
\end{figure}

For coordinate spheres with radii $R=11.28M_\odot$ to $R=16.93M_\odot$
in grid coordinates, the spins remain roughly constant in time.  The
different extraction spheres yield spins that agree to about 1\%, with
a consistent trend that larger extraction spheres result in slightly
larger spins (as already observed in the initial data).  The
horizontal dashed line in Fig.~\ref{fig:ChiVTZoomed} indicates the
spin measured on the stellar surface (i.e. not on a coordinate sphere)
in the initial data.  We thus find very good agreement between all
spin measurements, and conclude that the quasi-local spin is reliable
to about 1\%.

The extraction sphere $R=9.87M_\odot$ in Fig.~\ref{fig:ChiVTZoomed}
intersects the outer layers of the neutron star.  Because the
quasi-local spin captures only the angular momentum within the
extraction sphere, the value measured on $R=9.87M_\odot$ falls as our
comoving grid-coordinates cause this coordinate sphere to slowly move
deeper into the interior of the star.  This behavior, again, is
consistent with Fig.~\ref{fig:ChiVR}.

These tests of using multiple coordinate spheres were only run for
about half of the inspiral -- enough to establish that the method is
robust.  Subsequently, we report spins measured on the largest
coordinate sphere, $R=16.93M_\odot$.

The full behavior of the spin during the inspiral is shown in
figure~\ref{fig:ChiVTDifferentRes2} for both the {\tt S.4z} and {\tt S-.05z} runs.
Comparing the spin at different resolutions, we note that the data for
$k\!=\!1$ and $k\!=\!2$ are much closer to each other than compared to
$k\!=\!0$, indicating numerical convergence.  We note that the impact
of numerical resolution (as shown in
Fig.~\ref{fig:ChiVTDifferentRes2}) is small compared to the
uncertainty inherent from the choice of extraction sphere,
cf. Fig.~\ref{fig:ChiVTZoomed}.  We also note that for the first
$10000M_{\odot}$ of the run, the measured spin behaves as a constant,
as expected, albeit with some small oscillations. However,
afterward, we notice the
absolute value of the spin starts to decrease in both cases.
Finally, we note that in both cases, the spin measured in the inital
data on the stellar surface is within $\Delta \chi = 0.008$ of the spin
measured during the evolution.  

\begin{figure}
\includegraphics[width=0.95\columnwidth]{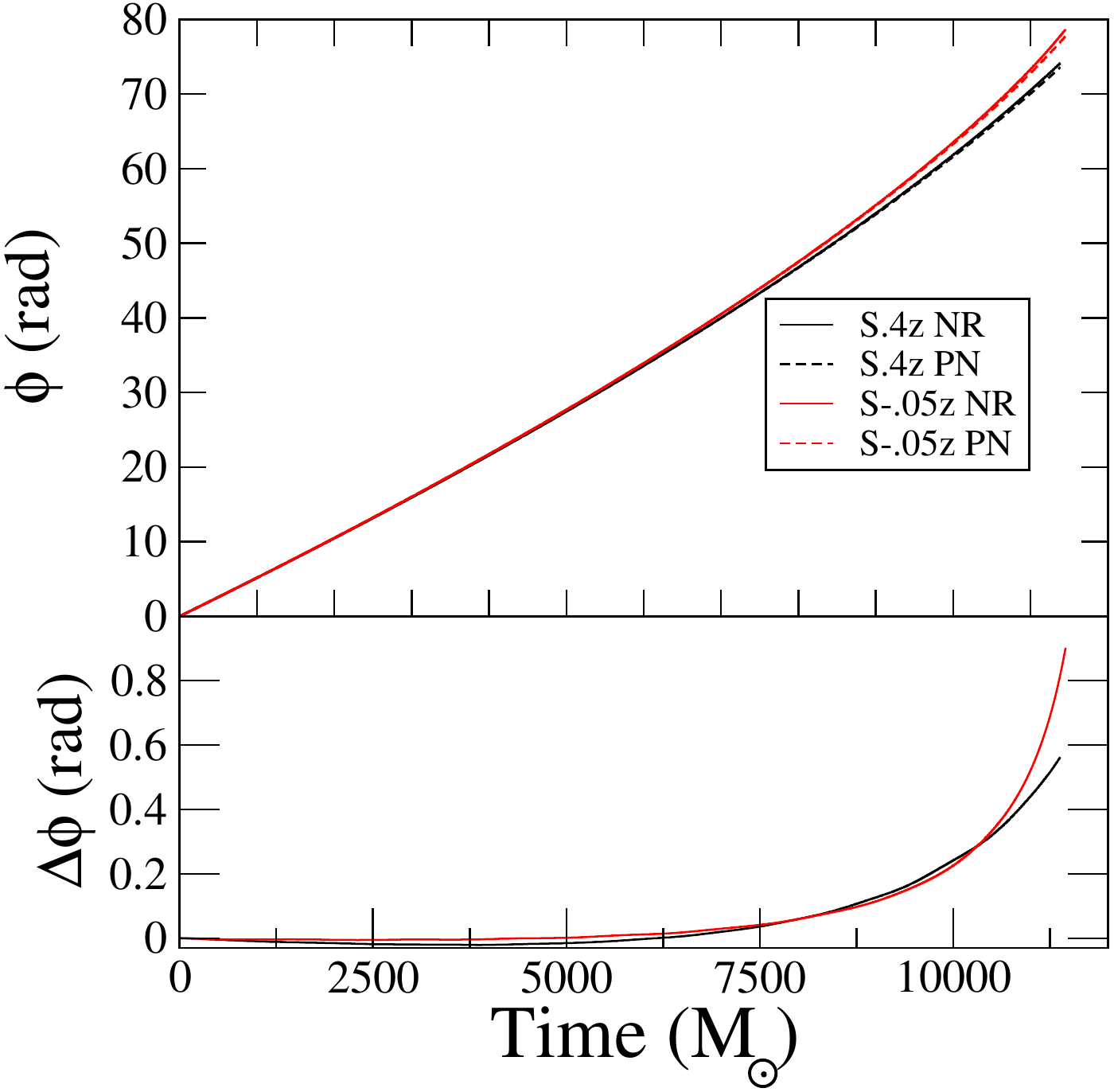}
\caption{ {\label{fig:Hangup}} Accumulated orbital phase as a function
  of time for our anti-aligned, {\tt S-.05z}, and aligned, {\tt S.4z}, runs. The
  dashed lines are Taylor T4 PN simulations. The PN simulations were
  matched to NR in the intervals [1109,3956] and [2090,4904]
  respectively. Qualitatively, there
  is excellent agreement with the numerical data. The lower panel
  shows the difference
  $\Delta\phi(t)=\phi_{\rm NR}(t)-\phi_{\rm PN}(t)$.}
\end{figure}

Finally, we compute the orbital phase 
\begin{equation}
\phi(t) = \int_0^t \Omega(t')\, dt',
\end{equation}

where the orbital frequency $\Omega(t)$ is given by
Eq.~(\ref{eq:Omega}).
The result is plotted in Fig.~\ref{fig:Hangup}, along with the

Post-Newtonian prediction for the same binary parameters (spins,
masses and initial orbital frequencies). We use the Taylor T4 model (see e.g.,\cite{Boyle2007}) at 3.5PN order expansion, with no tidal terms added, using
the matching techniques described in \cite{OssokineEtAl:2014}. We find
excellent qualitative agreement in both cases, thereby giving
additional evidence that our numerical simulations are working as
expected. We do find large late time growth in the phase difference,
however this is expected because we do not model tidal effects, which
become increasingly important at late times, in our
Post-Newtonian equations.

Figure~\ref{fig:AlignedGW} shows the gravitational waveforms for our
two non-precessing simulations. We extract the waves on a sphere of
radius $R=627M_{\odot}$.

\begin{figure}
\includegraphics[width=0.95\columnwidth]{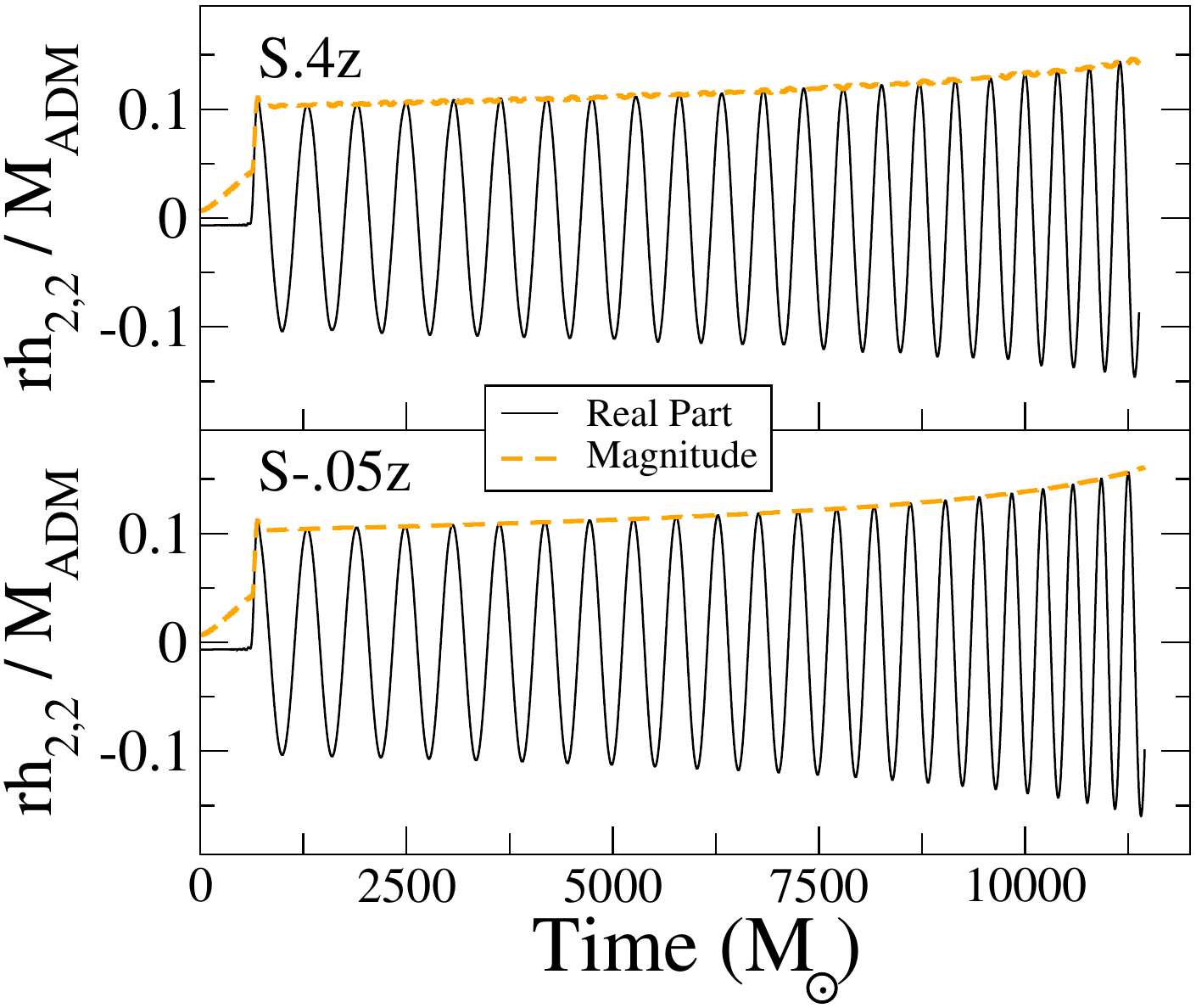}
\caption { {\label{fig:AlignedGW}} The gravitational waveforms for our
  anti-aligned, {\tt S-.05z}, and aligned, {\tt S.4z} runs. The black curve
  represents the real part of the waveform, $\Re(h_{2,2})$ while the
  orange curve represents the magnitude of the waveform.}
\end{figure}

\begin{figure}
\includegraphics[width=0.95\columnwidth]{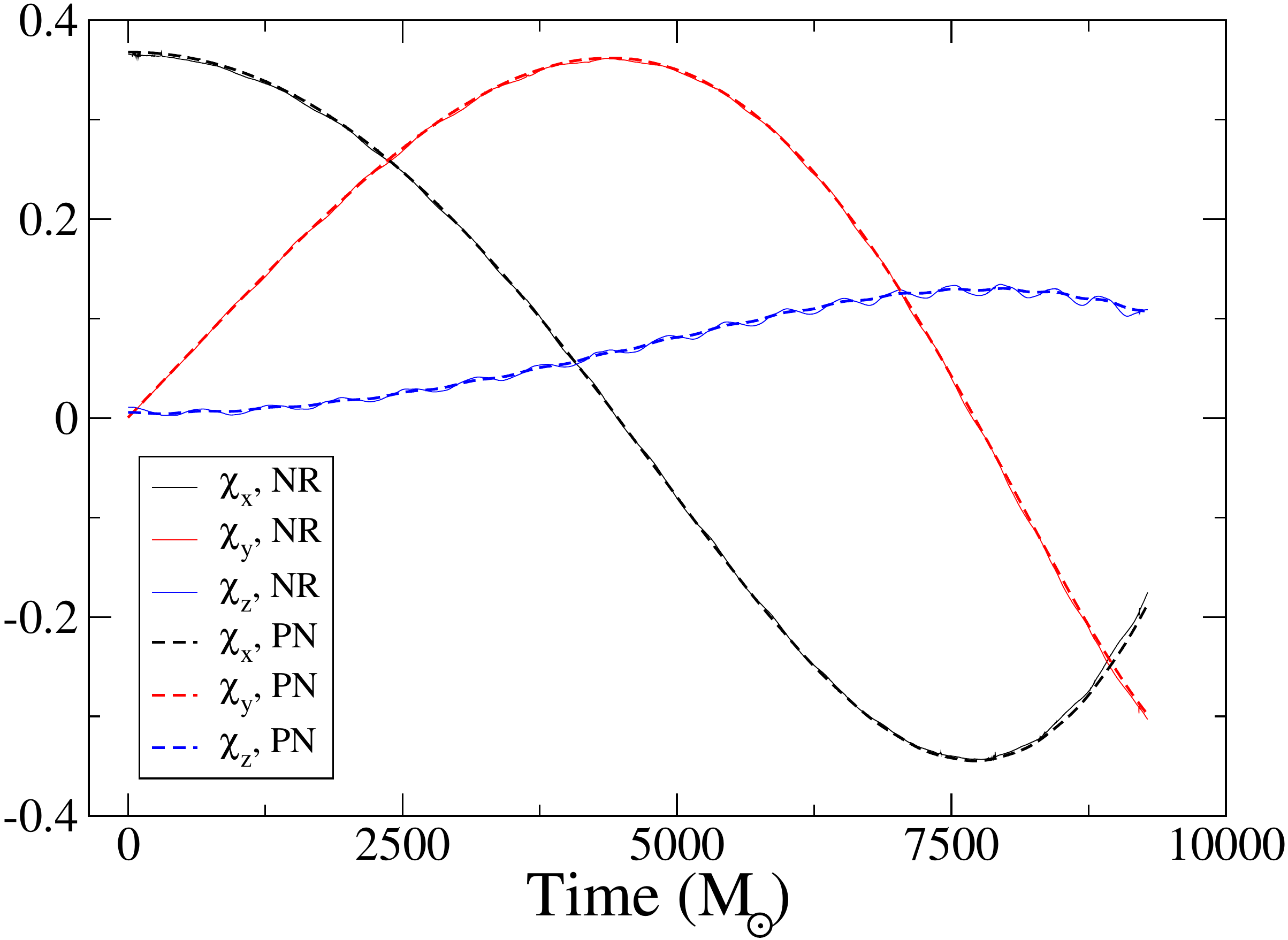}
  \caption{\label{fig:FullPrec} Spin-components of one of the neutron
    stars during the precessing simulation (thick, solid lines).  The
    dotted and dashed lines represent the unmatched and matched PN
    results respectively. The agreement between PN and NR is good for
    both PN simulations. The orbital frequency was evolved using the
    Taylor T4 approximant. The matching was done in the interval
    [1892,4575].}
\end{figure}

\subsection{Precession}

We now turn to the precessing simulation, {\tt S.4x}.
Figure~\ref{fig:FullPrec} shows the components of the spin-vector
$\vec\chi$ of one of the neutron stars, as a function of time.  The
quasi-local spin diagnostic returns a spin with nearly constant
magnitude, varying only by $\pm 0.002$ around its average value
$0.370$.  The spin components clearly precess, with the dominant
motion in the xy-plane (the initial orbital plane), with the
simulation completing about 2/3 of a precession cycle.  A z-component
of the NS spin also appears, indicating precession of the neutron star
spin out of the initial orbital plane.

\begin{figure}
\includegraphics[width=0.95\columnwidth]{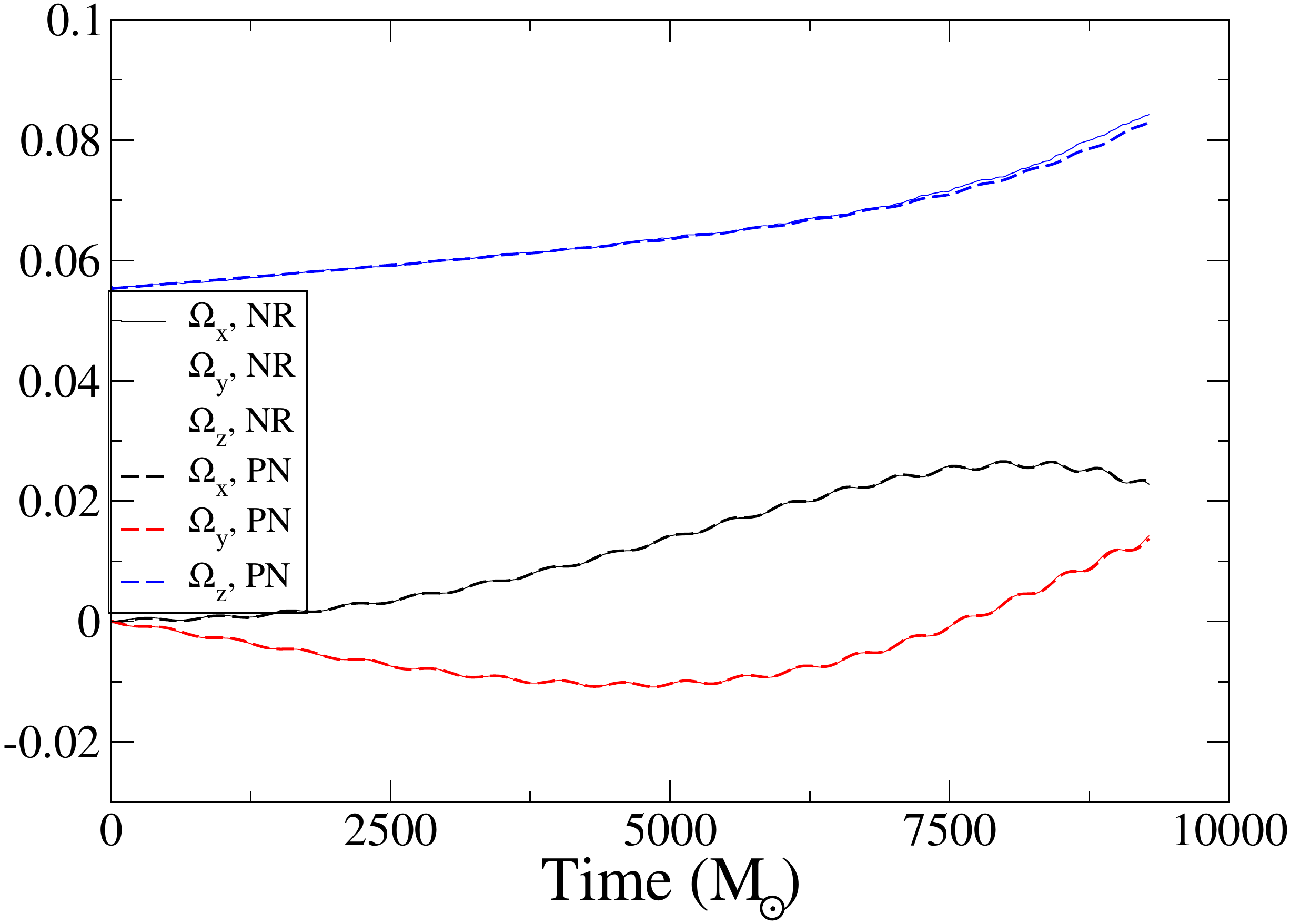}
\caption{\label{fig:OmegaVectorComparison} Components of the orbital
  frequency vector $\vec\Omega$.  Thick solid lines represent the
  precessing BNS simulation and thin
  dashed lines represent the matched  post-Newtonian simulations. The inclination reaches $\delta=0.34 {\rm rad}$ at $t=7600M_{\odot}$.}
\end{figure}

Fig.~\ref{fig:FullPrec} shows a comparison of spin precession
  between numerical relativity and Post-Newtonain theory.  We perform
  this comparison using
  the matching tecnhique in \cite{OssokineEtAl:2014}. This gives very good agreement between PN
  (dotted) and NR (solid) as shown by Fig.~\ref{fig:FullPrec}. The NS spins indeed precess as expected,
  thus confirming both the quality of quasi-local spin measures, as
  well as the performance of the PN equations. Note that z-component of the spin in the NR data undergoes oscillations that are unmodelled by PN. These occur on a timescale of half the orbital timescale. Similar effects were found in~\cite{OssokineEtAl:2014}. The origin of these oscillations remains unclear.   The precession of the
  orbital angular frequency is shown in
  Fig.~\ref{fig:OmegaVectorComparison}.  We find substantial
  precession away from the initial direction of the orbital frequency
  $\vec\Omega_0\propto \hat z$, with the angle $\delta$ between
  $\vec\Omega(t)$ and the z-axis reaching $20^\circ$.  Once again,
  the PN equations reproduce the precession
  features successfully. 

\begin{figure}
\includegraphics[width=0.9\columnwidth]{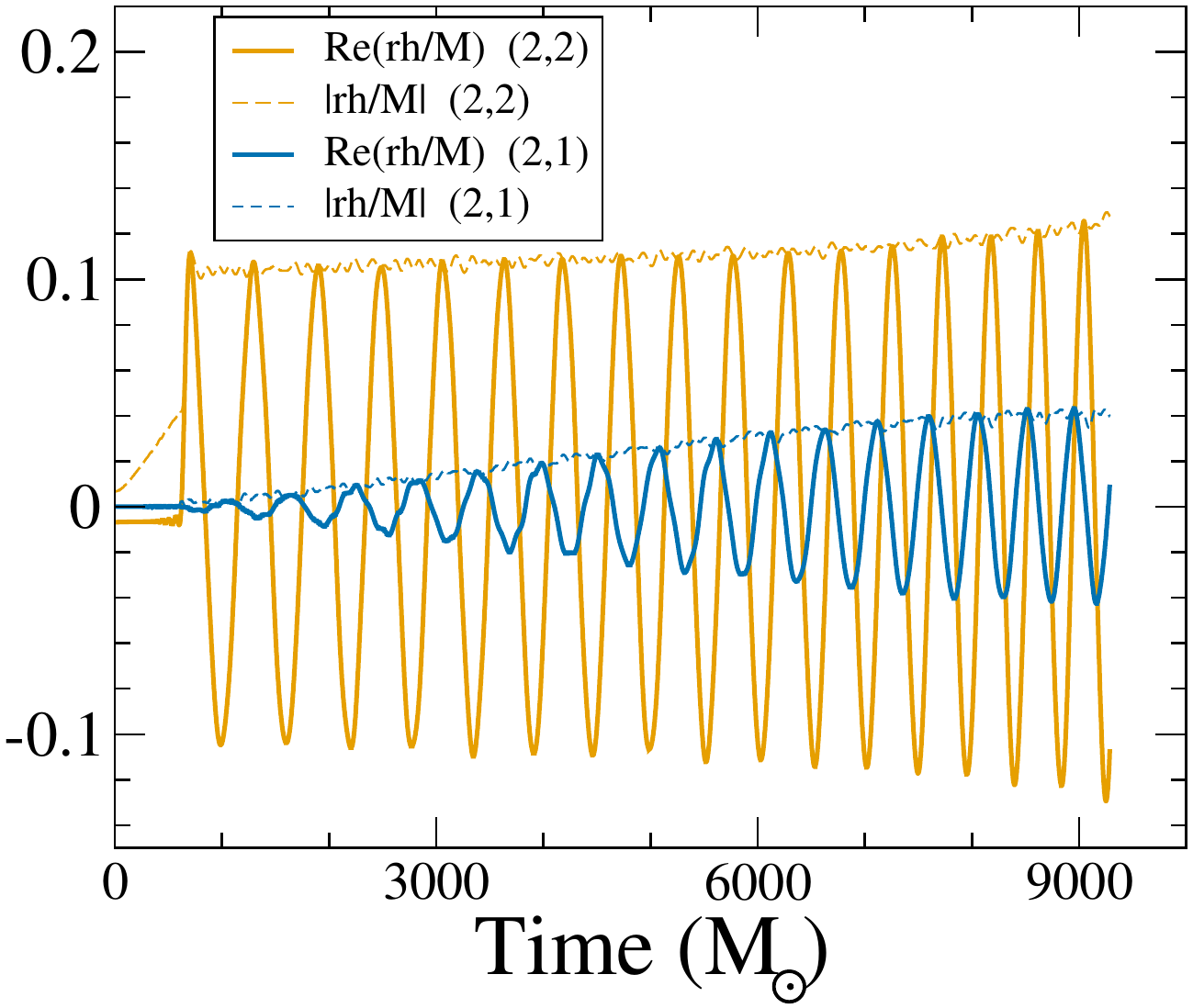}
\caption{\label{fig:PrecGW} Gravitational waveforms of our precessing
  run.  Shown are the $(l,m)=(2,2)$ and $(2,1)$ modes, as extracted in
  a spherical harmonic decomposition aligned with the z-axis.  The
  emergence of the (2,1) mode indicates precession of the orbital
  plane away from the xy-plane.}
\end{figure}

Finally, Fig.~\ref{fig:PrecGW} shows the (2,2) and the (2,1) spherical
harmonic modes of the gravitational wave-strain extracted at an
extraction surface of radius $R=647M_\odot$.  The
(l,m)=(2,1) mode would be identically zero for an equal-mass aligned
spin binary with orbital frequency parallel to the z-axis, so the
emergence of this mode once again indicates precession in this binary.

\subsection{Stellar Oscillations}
\label{sec:QNModes}

\begin{figure}
\includegraphics[width=0.95\columnwidth]{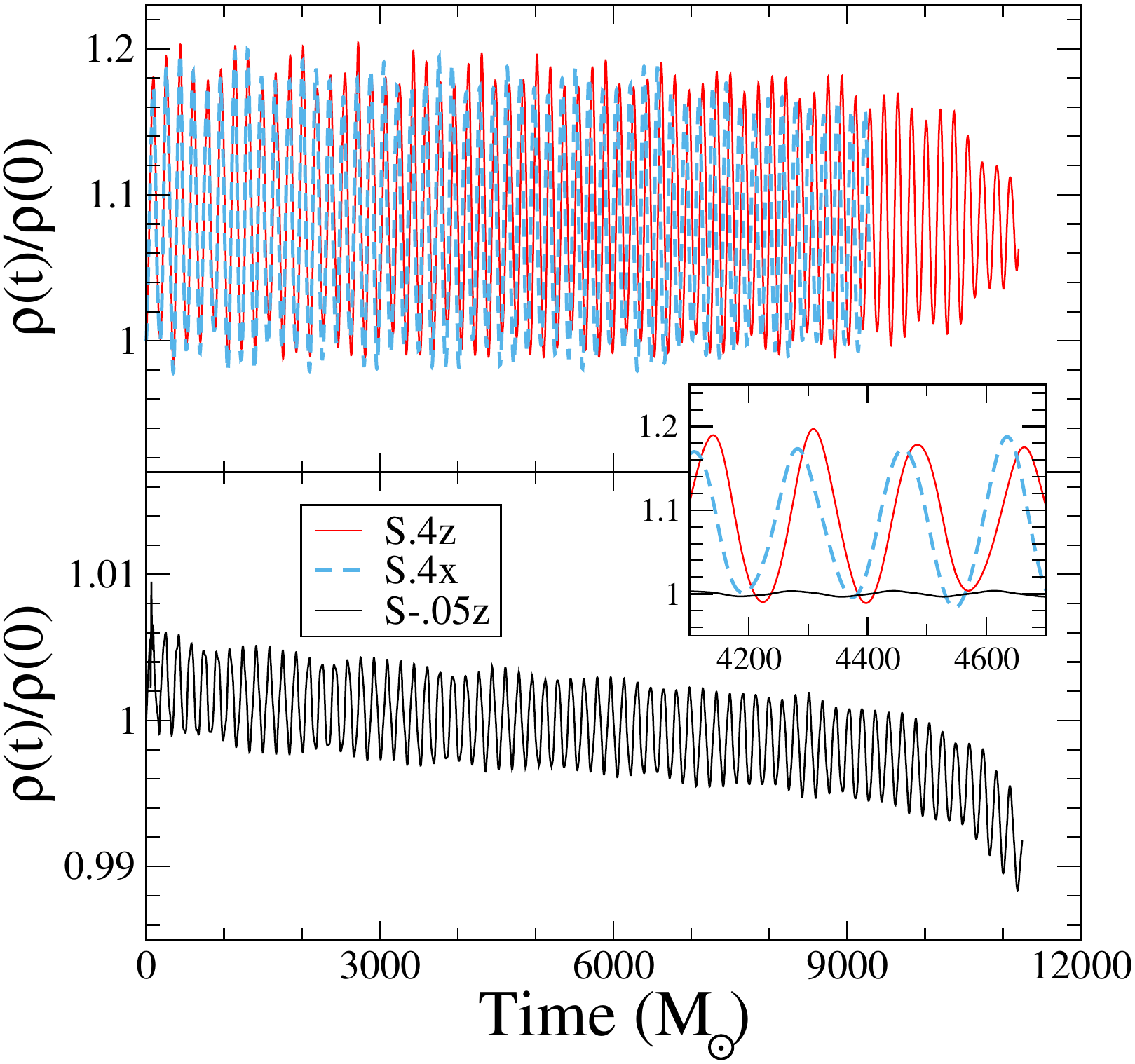}
\caption{\label{fig:RhoMax} The maximum density $\rho(t)$ in each of
  our runs, normalized by the initial maximum density $\rho(0)$.  The
  inset shows an enlargement of all three runs, illustrating that the
  oscillations are more pronounced in the high-spin simulations. }
\end{figure}

The rotating neutron stars constructed here show oscillations in the
central density, as plotted in Fig.~\ref{fig:RhoMax}.  In the low spin
run, the density oscillations have a peak-to-peak amplitude of about
0.6\%, whereas in the high-spin runs ({\tt S.4z} and {\tt S.4x}), the density
oscillations reach a peak-to-peak amplitude of 20\%.  The two
high-spin simulations show oscillations of nearly the same amplitude
and frequency, therefore oscillating nearly in phase throughout the
entire inspiral.  The oscillation-period is about
$177 M_{\odot} \sim 0.87\rm{ms}$, i.e. giving a frequency of
$1.15{\rm kHz}$.  It remains constant throughout the inspiral.  The low-spin
run {\tt S-0.5z} exhibits a slightly smaller oscillation period of about
$P\approx 170M_{\odot}\approx 0.84\rm{ms}$, i.e. a frequency of
$\approx 1.19{\rm kHz}$.

\begin{figure}
\includegraphics[width=0.95\columnwidth]{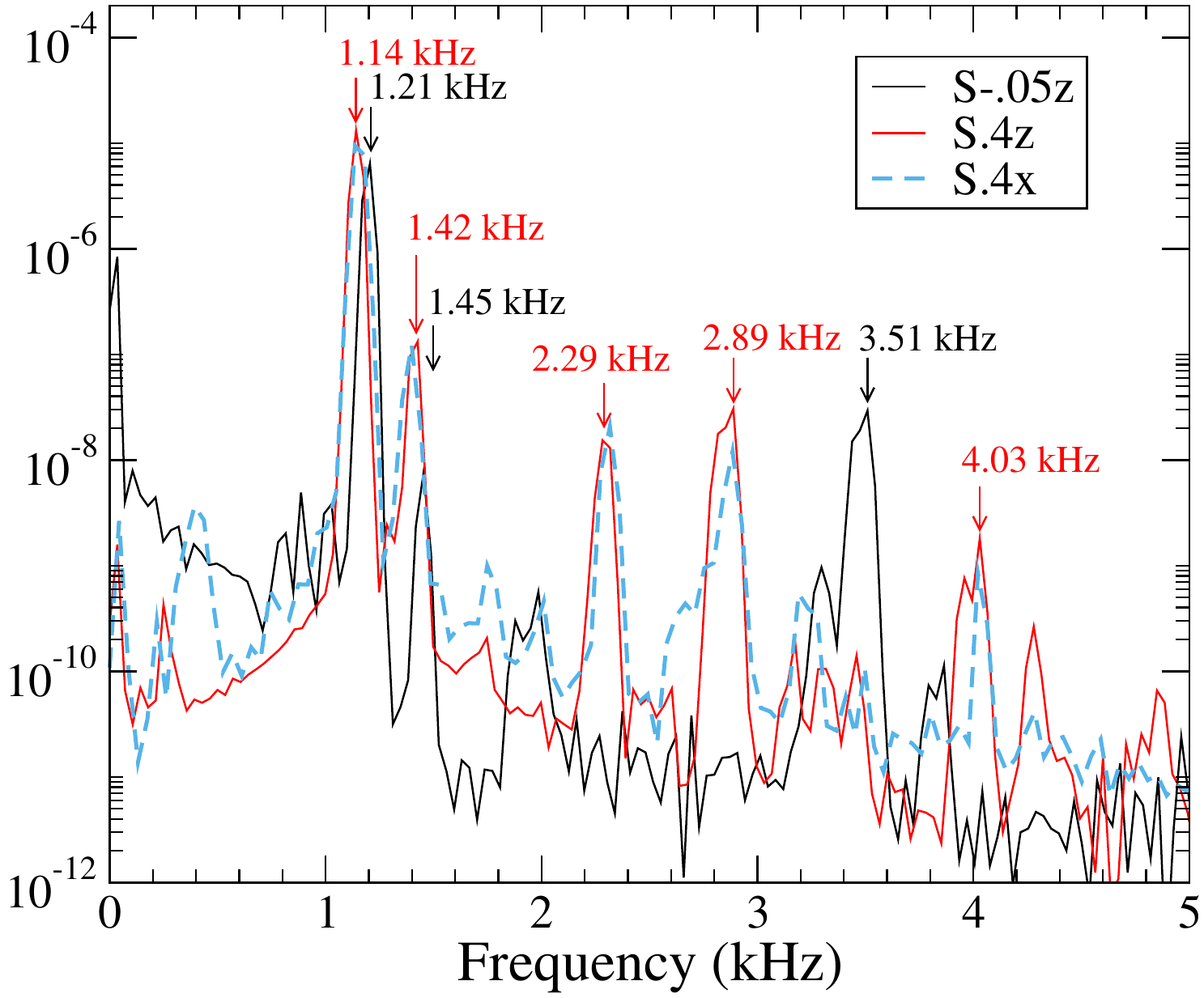}
\caption{\label{fig:Density_FFT} The Fourier transforms of the central
  density in all three of our runs. Labelled are the peak frequencies
  for the quasi-radial F mode and the $l=2$, $^2f$ mode.}
\end{figure}

To investigate the spectrum of the density oscillations, we perform a
Fourier-transform on $\rho(t)$.  The result is shown in
Fig.~\ref{fig:Density_FFT}.  The Fourier-transform confirms the
dominant frequencies just stated, and reveals several more frequency
components ranging up to 4kHz.  The high spin evolutions {\tt S.4z} and
{\tt S.4x} exhibit identical freqencies for all five discernible peaks.  In
contrast, the low-spin evolution {\tt S-.05z} shows different frequencies.  

We interpret these features as a collection of excited quasi-normal
modes in each neutron star.  The modes are excited because the initial
data is not precisely in equilibrium.  For the two high-spin cases the
neutron stars have similar spin, and therefore the same quasi-normal
modes, whereas in the low-spin model, the quasi-normal mode
frequencies differ due to the different magnitude of the spin.

To strengthen our interpretation, we consider the series of rotating,
relativistic, $\Gamma=2$ polytropes computed by Dimmelmeier et
al~\cite{Dimmelmeier:2005zk}. 

Ref.~\cite{Dimmelmeier:2005zk}'s model ``AU3'' has a central density of
$1.074 \times 10^{-3} M_{\odot}^{-2}$ and its rotation is
quantified through the ratio of polar to equatorial radius,
$r_p/r_e = 0.780$.
Meanwhile, our high-spin runs have a central density of
$1.02 \times 10^{-3} M_{\odot}^{-2}$ (measured as time-average of the data shown in
Fig.~\ref{fig:RhoMax}) and from our initial data, we find
$r_p/r_e \sim 0.8$. Given the similarity in these
values, we expect Ref.~\cite{Dimmelmeier:2005zk}'s ``AU3'' to approximate our high-spin
stars {\tt S.4x}, {\tt S.4z}.
Ref.~\cite{Dimmelmeier:2005zk} reports a frequency of
$f_F=1.283\rm{kHz}$ for the spherically symmetric ($\ell=0$) F-mode,
and a frequency $f_{2f}=1.537\rm{kHz}$ for the axisymmetric $\ell=2$
mode $^2f$.  These frequencies compare favorably with the two dominant
frequencies in Fig.~\ref{fig:Density_FFT}, $1.14\rm{kHz}$ and
$1.42\rm{kHz}$.

Presumably, the small differences in these frequencies can be
accounted for by the slight differences in stellar mass, radius, and
rotation.  Moreover, tidal interactions and orbital motion could
factor in, as well. In our figure~\ref{fig:Density_FFT} we also see
several other peaks at higher frequencies, which are reminiscent of
the overtones and mode couplings in figure 10 of
\cite{Dimmelmeier:2005zk}. If we identify our peak at
  $f_{H1}=4.03\rm{kHz}$ with the $H_1$ mode, then (in analogy to
  \cite{Dimmelmeier:2005zk} Fig.~10), 
  $f_{H1}-f_{F}=(4.03-1.14)\rm{kHz}=2.89\rm{kHz}$, and
  $2f_{F}=2.28\rm{kHz}$, two frequencies that are indeed present in
  our simulations.
Although we find clear indications of axisymmetric $\ell=2$-modes,
we note that their power is smaller by two orders of magnitude,
compared to the spherically symmetric, dominant $F$ mode. 

Turning to the low-spin run {\tt S.05z}, we note that if, to first order,
these frequencies scale like $f\sim\sqrt{\rho}$ (on dimensional
grounds), then we expect to see $F=1.22\rm{kHz}$ and
$^2f=1.49\rm{kHz}$. This is very close to what is seen.

The density oscillations discussed in this section are reflected in
analogous oscillations in various other diagnostic quantities, for
instance, the orbital frequency, Fig.~\ref{fig:OmegaDotComparison} and
the quasi-local spin as shown in Fig.~\ref{fig:ChiVTDifferentRes2}.
The dominant frequencies $1.14\rm{kHz}$ and $1.42\rm{kHz}$ can be
robustly identified throughout our data analysis. In
figure~\ref{fig:ManyQuantities} we plot the Fourier transform of the
density, the $(2,0)$ and $(2,2)$ gravitational wave strains, the
orbital angular velocity time derivative $d\Omega/dt$ and the
measured spin $\chi$ for the {\tt S.4z} run. All show peaks in power at
these two frequencies, $F\sim1.14\rm{kHz}$ and $^2f\sim1.4\rm{kHz}$.
Gold et. al. \cite{Gold:2011df}, in the simulation of close encounters in eccentric, irrotational, NSNS binaries,
find an excited f-mode frequency of 1.586 kHz.

\begin{figure}
\includegraphics[width=0.95\columnwidth]{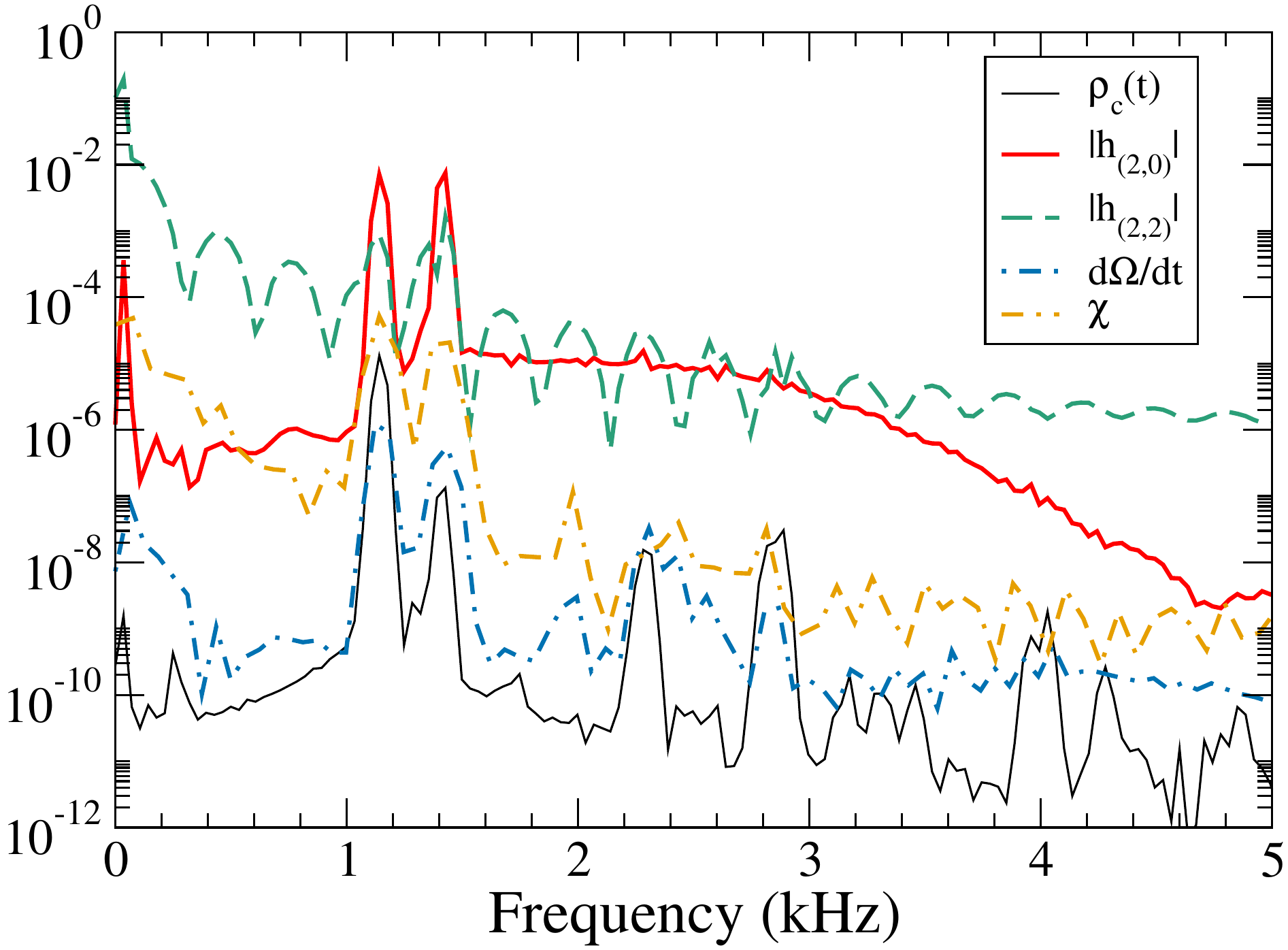}
\caption{\label{fig:ManyQuantities} Fourier transforms of the central
  density $\rho_c(t)$, two modes of the magnitude of gravitational wave strain
  $(|h_{2,2}|$ and $|h_{2,0}|$), $\dot{\Omega}$ and $\chi$ for the {\tt S.4z}
  run.  All quantities show excess power at $1.14\rm{kHz}$ and
  $1.4\rm{kHz}$, corresponding to the frequencies of excited neutron
  star quasi-normal modes. 
}
\end{figure}

We believe  that the  stellar  modes are  excited because  the
  initial  data  are  not  in  perfect  equilibrium.   We  expect  the
  quasi-equilibrium  approximations   that  enter  the   initial  data
  formalism to become less valid  at higher spins, consistent with our
  observation that the high spin models exhibit stronger oscillations.
  This  interpretation is  strengthened by  additional simulations  of
  neutron stars at larger separation. Increasing   the  intial
  separation  by  a  factor  1.5,  while  keeping  the  same  rotation
  parameter  $\omega$  as  in  the  {\tt S.4z}-case,  we  find  quasi-normal
  oscillations   of  similar   amplitude  than   in  {\tt S.4z}.    If  the
  oscillations were  caused by  the neglect  of tidal  deformation, we
  would expect the amplitude to drop with the 3rd power of separation,
  inconsistent with our results.

  Finally, we point out that the radial rotation profile,
  cf. Eq.~(\ref{eq:UniformRotation}) influences the amplitude of the
  induced quasi-normal oscillations.  If the initial data is
  constructed with the rotation profile
  Eq.~(\ref{eq:ConformalUniformRotation}), instead of
  equation~\ref{eq:UniformRotation}, then the amplitude of the density oscillations for
  high spin doubles. This further supports our conjecture that the
  origin of this mode comes from non-equilibrium initial data.

\section{Discussion}
\label{sec:Discussion}

In this paper we implement Tichy's method~\cite{Tichy:2012rp} to
construct binary neutron star initial data with arbitrary rotation
rates.  We demonstrate that our implementation is exponentially
convergent, as expected for the employed spectral methods.

We measure the spin of the resulting neutron stars using the
quasi-local angular momentum
formalism~\cite{BrownYork1993,Cook2007,Lovelace2008,OwenThesis}.  The
resulting angular momentum is found to be nearly independent on the
precise choice of extraction sphere, cf. Fig.~\ref{fig:ChiVR}, and
provides a means to define the quasi-local angular momentum of each
neutron star to about 1\%, both in the initial data and during the
evolution, cf. Fig.~\ref{fig:ChiVTZoomed}.  We are able to construct
binary neutron star initial data with dimensionless angular momentum
of each star as large as $\chi=S/M^2\sim 0.43$, both for the case of
aligned spins, and also for a precessing binary where the initial
neutron star spins are tangential to the initial orbital plane.

For irrotational BNS initial data sets, we find a quasi-local
  angular momentum of $\chi\sim 2\times 10^{-4}$,
  cf. Fig.~\ref{fig:ChiVOmega}. This spin is small enough that present
  waveform modeling studies for BNS
  (e.g.~\cite{Bernuzzi:2014owa,Baiotti2011,Baiotti:2010xh}) are not
  yet limited by residual spin.

When evolving the initial data sets, the dimensionless spin measured
in the initial data drops by about 0.004, and then remains constant
through the 10 inspiral orbits for which we evolved the neutron star
binaries.  During these evolutions, we also demonstrated iterative
eccentricity removal: By analyzing the orbital frequency $\Omega(t)$
during the first few orbits, we can correct the initial data
parameters $\Omega_0$ and $\dot a_0$, and thus decrease the orbital
eccentricity from $e\approx 0.01$ to $e\lesssim 0.001$.

For the precessing simulation {\tt S.4x}, we find precession of the neutron
star spin directions.  The numerically established precession of the
spin axes and of the orbital angular momentum agrees well with
post-Newtonian predictions.

The rotating neutron stars constructed here exhibit clear signals of
exciting quasi-normal modes.  We are able to identify multiple modes
in the Fourier spectrum of the central density.  The amplitude of the
excited quasi-normal modes increases steeply with rotation rate of the
neutron stars.  For {\tt S-.05z} (spin magnitude $\chi=0.045$) the density
oscillations have peak-to-peak amplitude of 0.6\%, raising to 20\% for
the two runs with high spins ({\tt S.4x} and {\tt S.4z}).\\

\begin{acknowledgments}

We thank Rob Owen and Geoffrey Lovelace for discus- sions on
quasi-local spins.  Calculations were performed with the Spectral
Einstein Code ({\tt SpEC})~\cite{SpECwebsite}.  We gratefully
acknowledge support for this research at CITA from NSERC of Canada,
the Canada Research Chairs Program, the Canadian Institute for
Advanced Research, and the Vincent and Beatrice Tremaine Postdoctoral
Fellowship (F.F.); at LBNL from NASA through Einstein Postdoctoral
Fellowship grant PF4-150122 (F.F.) awarded by the Chandra X-ray
Center, which is operated by the Smithsonian Astrophysical Observatory
for NASA under contract NAS8-03060; at Caltech from the Sherman
Fairchild Foundation and NSF grants PHY-1440083, PHY-1404569,
PHY-1068881, CAREER PHY-1151197, TCAN AST-1333520, and NASA ATP grant
NNX11AC37G; at Cornell from the Sherman Fairchild Foundation and NSF
Grants PHY-1306125 and AST-1333129; and at WSU from NSF Grant
PHY-1402916.  Calculations were performed at the GPC supercomputer at
the SciNet HPC Consortium~\cite{scinet}; SciNet is funded by: the
Canada Foundation for Innovation (CFI) under the auspices of Compute
Canada; the Government of Ontario; Ontario Research Fund (ORF) --
Research Excellence; and the University of Toronto. Further
calculations were performed on the Briar\'ee cluster at Sherbrooke
University, managed by Calcul Qu\'ebec and Compute Canada and with
operation funded by the Canada Foundation for Innovation (CFI),
Minist\'ere de l'\'Economie, de l'Innovation et des Exportations du
Quebec (MEIE), RMGA and the Fonds de recherche du Qu\'ebec - Nature et
Technologies (FRQ-NT); on the Zwicky cluster at Caltech, which is
supported by the Sherman Fairchild Foundation and by NSF award
PHY-0960291; on the NSF XSEDE network under grant TG-PHY990007N; on
the NSF/NCSA Blue Waters at the University of Illinois with allocation
jr6 under NSF PRAC Award ACI-1440083.
\end{acknowledgments}

\newpage
\bibliography{../References/References}

\end{document}